\documentclass[
reprint, 
amsmath,amssymb,
aps,
prl,
floatfix,
longbibliography,
]{revtex4-2}

\usepackage{graphicx}
\usepackage{bm}
\usepackage{physics}
\usepackage[usenames, dvipsnames]{color}
\usepackage{textgreek}

\definecolor{NewBlue}{rgb}{0, 0, 0.41}
\definecolor{NewRed}{rgb}{0.6, 0.07, 0.07}
\usepackage[colorlinks,
    linkcolor=NewBlue,
    citecolor=NewBlue,
    urlcolor=NewRed]{hyperref}

\begin{document}

\title{Control of individual electron-spin pairs in an electron-spin bath}

\author{H. P. Bartling$^{1,2}$}
\thanks{These authors contributed equally to this work.}
\author{N. Demetriou$^{1,2}$}
\thanks{These authors contributed equally to this work.}
\author{N. C. F. Zutt$^{1,2}$}
\author{D. Kwiatkowski$^{1,2}$}
\author{M. J. Degen$^{1,2}$}
\author{\\S. J. H. Loenen$^{1,2}$}
\author{C. E. Bradley$^{1,2}$}
\author{M. Markham$^3$} 
\author{D. J. Twitchen$^3$}
\author{T. H. Taminiau$^{1,2}$}
\email{T.H.Taminiau@TUDelft.nl}

\affiliation{$^{1}$QuTech, Delft University of Technology, PO Box 5046, 2600 GA Delft, The Netherlands}
\affiliation{$^{2}$Kavli Institute of Nanoscience Delft, Delft University of Technology, PO Box 5046, 2600 GA Delft, The Netherlands}
\affiliation{$^{3}$Element Six, Fermi Avenue, Harwell Oxford, Didcot, Oxfordshire, OX11 0QR, United Kingdom}

\date{\today}

\begin{abstract} 

The decoherence of a central electron spin due to the dynamics of a coupled electron-spin bath is a core problem in solid-state spin physics. Ensemble experiments have studied the central spin coherence in detail, but such experiments average out the underlying quantum dynamics of the bath. Here, we show the coherent back-action of an individual NV center on an electron-spin bath and use it to detect, prepare and control the dynamics of a pair of bath spins. We image the NV-pair system with sub-nanometer resolution and reveal a long dephasing time ($T_2^* = 44(9)$ ms) for a qubit encoded in the electron-spin pair. Our experiment reveals the microscopic quantum dynamics that underlie the central spin decoherence and provides new opportunities for controlling and sensing interacting spin systems.

\end{abstract}

\maketitle

Solid-state spins provide a versatile platform for quantum science and technology, as well as for studying the fundamentals of spin coherence. A canonical case is the central spin problem: a single, central spin coupled to a surrounding bath of interacting spins \cite{Klauder1962,hanson2008,Witzel2010,deLange2010,Witzel2012,Zhao2012Decoherence,Witzel2014,Seo2016, Bauch2020, Park2022}. A common approach to protect the central spin from decoherence due to the bath spins is to use echo or decoupling sequences. Under such echo sequences, the system undergoes complex quantum dynamics that depend on the microscopic bath configuration and on the back-action of the central spin on the interacting spin bath \cite{Klauder1962,hanson2008,Witzel2010,deLange2010,Witzel2012,Zhao2012Decoherence,Witzel2014,Seo2016, Bauch2020, Park2022}. 

For an electron spin in a nuclear-spin bath, the large magnetic moment of the electron spin strongly affects the nuclear-spin bath evolution. The resulting back-action creates rich dynamics under echo sequences on the central spin \cite{Taminiau2014,Shi2013,Abobeih2018} and has enabled the control of tens of individual nuclear spins \cite{Bradley2019,Vandestolpe2023}, pairs of coupled nuclear spins \cite{Bartling2022, Abobeih2018,Zhao2011,Shi2013}, and collective excitations \cite{Jackson2021,Ruskuc2022} in the spin bath. 

For an electron spin in an electron-spin bath, the effect of back-action is more subtle. All couplings are of similar strength and they are typically weak compared to the energy splittings. The resulting central spin decoherence has been investigated in detail in ensemble experiments \cite{Witzel2010,Tyryshkin2012,Witzel2012,Park2022,Bauch2020}, which have been described by semi-classical models, as well as by fully quantum models using approximate numerical methods \cite{Park2022,Ye2019,Seo2016,Witzel2010,Witzel2012,Bauch2020}. In such ensemble experiments, the underlying microscopic quantum dynamics are averaged out. Single-spin experiments have been performed with NV centers in diamond \cite{hanson2008,deLange2010, DeLange2012,Knowles2014, Degen2021, Rosenfeld2018, Cooper2020,Knowles2016,Yudilevich2022}. However, the coherence under echo sequences \footnote{Note that double-resonance sequences, which manipulate multiple spins, have been exploited to detect single electron spins \cite{Degen2021, Cooper2020, Knowles2016, Yudilevich2022}, including P1 centers, as well as signatures of flip-flop dynamics in spin clusters \cite{Rosenfeld2018}} could be satisfactorily described by an effective magnetic field noise (an Ornstein-Uhlenbeck process) \cite{deLange2010, Klauder1962, Witzel2014, Bauch2020}. In this classical model of the spin bath, the back-action of the central spin is neglected and the central limit theorem is used to approximate the bath as Gaussian, forgoing the microscopic quantum dynamics. 

Here, we show that the microscopic flip-flop dynamics in an electron-spin bath can be experimentally accessed and controlled using echo sequences on a central spin. Compared to previous work, we observe a single central electron spin, rather than an ensemble average, and use time-resolved correlations and real-time logic to prepare and observe specific configurations of the bath, rather than averaging over all states. We demonstrate strong back-action of a central electron spin on the dynamics of an individual spin pair in the bath, and use this coherent interaction to detect, image and control the spin pair. We show that the spin pair can be used to encode a controllable qubit with long coherence times ($T_2^* = 44(9)$ ms), due to a combination of a decoherence-free subspace and a clock transition. Our results directly access the microscopic quantum dynamics that underlie the central spin decoherence and provide new opportunities for controlling interacting spin systems.

\begin{figure}
\includegraphics[width = \columnwidth]{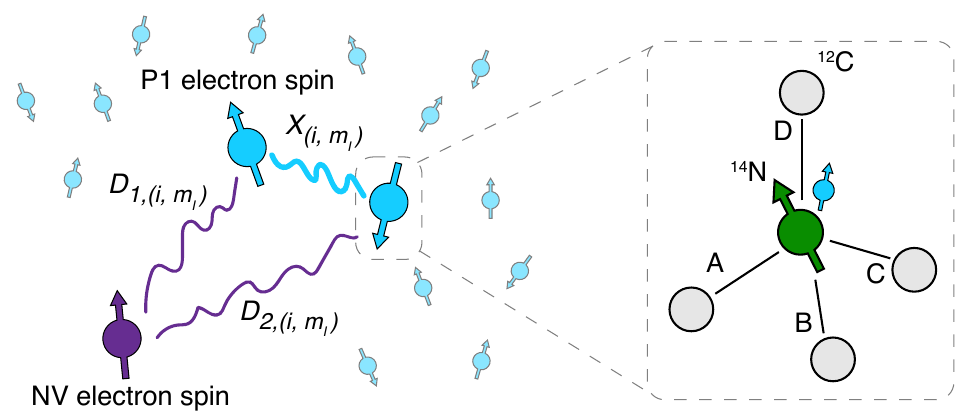}
\caption{\textbf{Schematic of the spin system.} We detect and control the dynamics of a pair of P1 electron spins in a P1 bath through the back-action from an NV center. The inset shows the lattice structure of a P1 center in diamond with the $^{14}$N nuclear spin (spin states $m_I \in \{-1,0,+1\}$) and four Jahn-Teller axes ($i \in \{A, B, C,D\}$). $X_{(i,m_I)}$ is the effective coupling between the P1 electron spins and $D_{(i,m_I)}$ are the effective couplings with the NV center electron spin.}
\label{fig1}
\end{figure}

We investigate a single nitrogen-vacancy (NV) center in diamond surrounded by a bath of P1 centers (nitrogen defects) at a temperature of 3.3 K (Fig. \ref{fig1}). The P1 concentration is $\sim 75$ ppb, and the estimated $^{13}$C concentration is $0.01\%$ \cite{Degen2021,Bradley2022}. The NV electron spin acts as the central spin and is initialized optically and read out using spin-selective optical excitation (637 nm) \cite{Degen2021,Bartling2022}. 

The P1 centers have multiple internal, dynamic degrees of freedom: the electron spin-1/2, four different Jahn-Teller (JT) axes and a spin-1 $^{14}$N nuclear spin (Fig. \ref{fig1}). The P1 Hamiltonian for JT axis $i \in \{A, B, C, D\}$ is \cite{Smith1959}
\begin{equation}
    H_{P1,i} = \gamma_e \mathbf{B} \cdot \mathbf{J} + \gamma_n \mathbf{B} \cdot \mathbf{I} + \mathbf{J} \cdot \mathbf{{A}}_i \cdot \mathbf{I} + \mathbf{I} \cdot \mathbf{{P}}_i \cdot \mathbf{I}.
\end{equation}
where $\gamma_e$ ($\gamma_n$) is the electron (nitrogen) gyromagnetic ratio and $\mathbf{J}$ ($\mathbf{I}$) is the electron spin-1/2 ($^{14}$N nuclear spin-1) operator vector. $\mathbf{{A}}_i$ ($\mathbf{{P}}_i$) is the hyperfine (quadrupole) tensor where the subscript $i$ indicates the Jahn-Teller axis \cite{Cook1966,Degen2021}. We apply a few-degree misaligned magnetic field with respect to the NV axis $\mathbf{B} = [2.43(2),1.42(3),45.552(3)]$ G to lift the degeneracy for the different JT axes (Supplementary Sec. VI).

Since the NV-P1 dipolar coupling can be approximated as $\hat{S}_z \hat{J}_z$ \footnote{In contrast, for a nuclear-spin bath the $\hat{S}_z \hat{J}_x$ terms are typically important \cite{Taminiau2014,Zhao2012Decoherence}}, echo sequences on the NV electron spin primarily probe the energy-conserving flip-flop dynamics of the P1 bath due to the dipolar P1-P1 couplings. Whether flip-flops between two P1 centers are allowed depends on their electron and $^{14}$N spin states, on their JT axes, and on the local magnetic field due to the NV, $^{13}$C spins, and other nearby P1 centers. Therefore, the dynamics are complex, depend strongly on the specific microscopic configuration, and change over time.

We probe and prepare specific bath configurations by performing time-resolved experiments through repeated NV measurement sequences. This is made possible by the long lifetime of the JT axis and $^{14}$N spin at cryogenic temperatures and by a low-intensity resonant readout of the NV spin that only weakly perturbs the P1 center states \cite{Degen2021}. Previous room-temperature experiments with high-power off-resonant lasers rapidly average over all P1 states \cite{hanson2008, deLange2010, DeLange2012,Ammerlaan1981,Heremans2009}.

\begin{figure}
\includegraphics[width = \columnwidth]{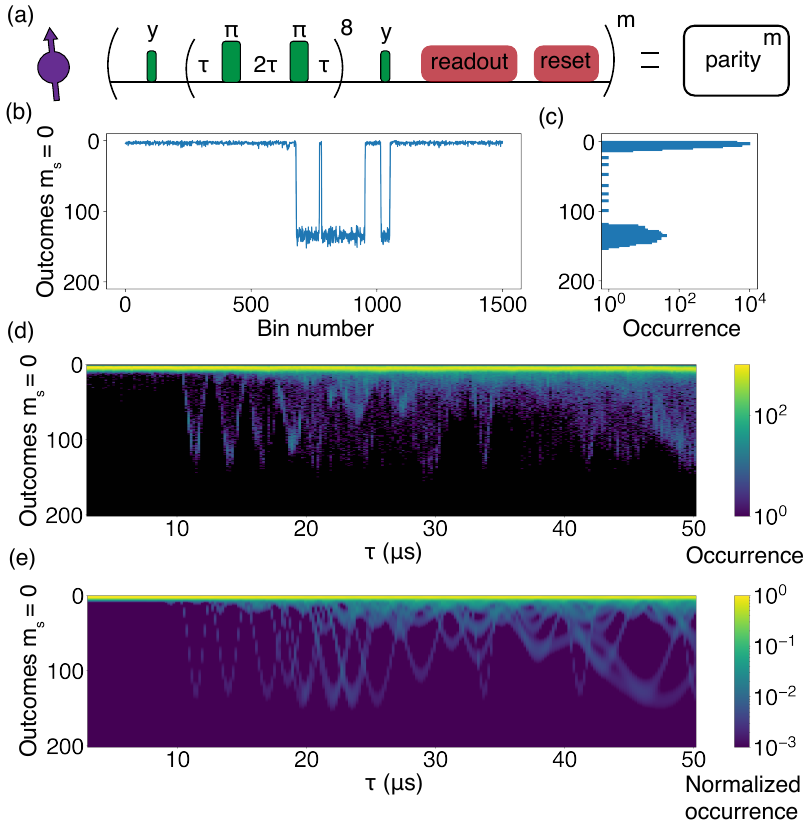}
\caption{\textbf{Repetitive dynamical decoupling spectroscopy of a P1 center bath.} \textbf{(a)} Experimental sequence. \textbf{(b)} Time trace for $\tau = 14.2$ \textmugreek s and bin-size $m = 200$. \textbf{(c)} Histogram of a 3-minute-long time trace for $\tau = 14.2$ \textmugreek s and $m = 200$. We rarely observe a high number of $m_s = 0$ occurrences ($\sim 1.3 \%$). Due to limited observations, the fraction of $\sim 1.3 \%$ is likely not a precise measure of the probability of occurrence. \textbf{(d)} Repetitive dynamical decoupling spectroscopy of a P1 bath surrounding an NV center. We apply the sequence shown in (a) for $m = 200$. \textbf{(e)} Simulation of the repetitive dynamical decoupling spectroscopy in (d) for a system of one NV center and two P1 centers with the positions as obtained in Fig. \ref{fig4}.}
\label{fig2}
\end{figure}

We apply dynamical decoupling sequences consisting of $\pi$-pulses with variable spacing $2\tau$ (Fig. \ref{fig2}a), which sense the bath dynamics around a frequency of $1/(4\tau)$. We repeatedly apply the sequence, bin $m$ outcomes together and analyze the signal and correlations over time. Figure \ref{fig2}b shows a time trace, revealing discrete jumps in the NV coherence. A longer-time histogram (Fig. \ref{fig2}c) reveals that the signal is a rare occurrence, which would be easily lost in the noise in a time-averaged measurement.
 
We create a map of the bath dynamics by collecting histograms as a function of the interpulse delay $\tau$ (Fig. \ref{fig2}d). The result shows distinct resonances for various values of $\tau$, which we attribute to two coupled P1 centers in the bath switching to different electron-spin, JT and $^{14}$N configurations. For each configuration, the P1 spin pair flip-flops with a characteristic frequency, which is resonant with the sensing sequence for a particular $\tau$. 

To analyze the results, we consider a single pair of P1 centers. For a large magnetic field, the electron- and nuclear-spin basis states are proper P1 eigenstates. Energy-conserving electron-spin flip-flops are then allowed when the two P1 centers have identical JT and $^{14}$N states. As exploited extensively for nuclear-spin pairs \cite{Zhao2011,Shi2013,Abobeih2018,Bartling2022}, the dynamics can then be described by a pseudo-spin in the anti-parallel spin subspace ($\ket{\Uparrow} = \ket{\uparrow \downarrow}$ and $\ket{\Downarrow} = \ket{\downarrow \uparrow}$). 

The pseudo-spin Hamiltonian \cite{Zhao2011,Shi2013,Abobeih2018,Bartling2022}, including the effect of the NV center, is:
\begin{equation}
    H_{(i,m_I)} = X_{(i,m_I)} \hat{S}_x + m_s Z_{(i,m_I)} \hat{S}_z, \label{pseudo-spin H}
\end{equation}
where $\hat{S}_x, \hat{S}_z$ are spin-1/2 operators, $X_{(i,m_I)}$ is the effective spin-pair coupling, $Z_{(i,m_I)}$ is a detuning due to the different couplings to the central NV spin, and $m_s$ is the NV spin projection. For large magnetic fields, there are 12 such Hamiltonians (four JT axes and three $^{14}$N states), all with equal values for $X$ and $Z$ (Supplementary Sec. II). For the field applied here, which is of the order of the hyperfine interaction ($\gamma_e B \sim A_\parallel, A_\perp$), the electron- and nuclear-spin states mix. Therefore, flip-flop interactions involving the nuclear spin are possible, and $X_{(i,m_I)}$ and $Z_{(i,m_I)}$ depend on the JT and spin states involved. We use the high-field spin labels for simplicity, but take the modified eigenstates and additional flip-flop interactions into account in our analysis.

Next, we demonstrate the initialization, control and measurement of the P1-pair state. From the mathematical equivalence with previous work \cite{ Zhao2012Decoherence,Taminiau2014,Bartling2022}, it follows that the Hamiltonian in Equation \ref{pseudo-spin H} yields an effective $\hat{S}_z^{\text{NV}} \hat{S}_z$ interaction under a resonant dynamical decoupling sequence with $2\tau = \pi/\omega_r$ with $\omega_r = \sqrt{X^2+(Z/2)^2}$. The NV electron spin thus picks up a positive or negative phase depending on the state of the P1-pair pseudo-spin \cite{Bartling2022}. No phase is picked up when the pair is in the parallel subspace ($\ket{\uparrow \uparrow}$, $\ket{\downarrow \downarrow}$), nor for any combination of JT and $^{14}$N states that do not cause flip-flop dynamics at the resonant frequency $\omega_r$.

To initialize the P1 pair in a particular JT and $^{14}$N state and in the anti-parallel subspace, we apply repeated `parity' readouts (Fig. \ref{fig2}a), and put a threshold on the obtained counts. We implement real-time logic to speed up the initialization procedure: during the $50$ parity readouts we keep track of the obtained counts and we restart the procedure if heralding successful preparation becomes unlikely (Supplementary Sec. XI). This yields a $\sim 10$x speed-up of the experiments and is essential for enabling the presented measurements.

\begin{figure}
\includegraphics[width = \columnwidth]{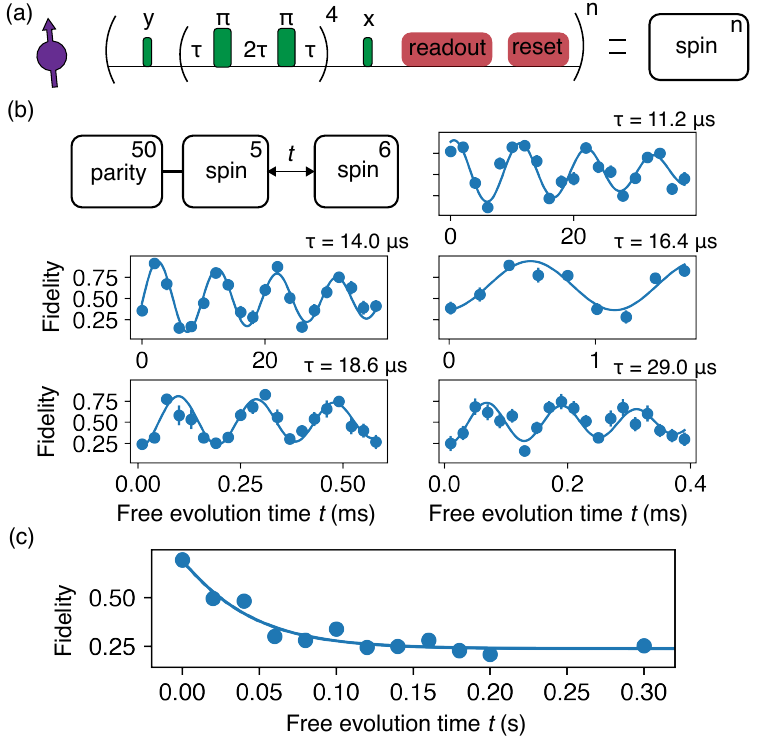}
\caption{\textbf{Ramsey measurements and coherence of the P1 spin pair.} \textbf{(a)} Experimental sequence to measure the pseudo-spin state ($\{\ket{\uparrow \downarrow}\}$ vs. $\{\ket{\downarrow \uparrow}\}$) (Supplementary Sec. III). \textbf{(b)} Ramsey measurements for five interpulse delays $\tau$ chosen at signal dips in Fig. \ref{fig2}d. We apply $50$ parity readouts and herald initialization for $\ge 15$ counts. We then apply 5 spin readouts, which herald initialization in $\ket{\uparrow \downarrow}$ ($\ket{\downarrow \uparrow}$) for $\ge 1$ ($=0$) counts. We use $6$ spin readouts to measure the final spin-pair state, assigning $\ge 2$ ($\le 1$) counts to $\ket{\uparrow \downarrow}$ ($\ket{\downarrow \uparrow}$). The contrast is limited by the pseudo-spin dephasing during the spin initialization and readout (Supplementary Sec. XII). \textbf{(c)} Bloch vector length measurement of the spin pair at $\tau = 14.0$ \textmugreek s, obtaining $T_2^* = 44(9)$ ms. The data is not corrected for the readout infidelity.}
\label{fig3}
\end{figure}

To initialize the spin pair, we apply repeated `spin' readouts (Fig. \ref{fig3}a). Subsequent spin measurements are time-matched to account for the evolution of the spin pair during one spin readout, similar to previous experiments with repeated measurements on precessing nuclear spins \cite{Bartling2022,Cujia2019} (Supplementary Sec. XII). 

By choosing a different interpulse delay $\tau$, we can address different JT and $^{14}$N states. We can thus measure the dependence of the electron-electron couplings ($X$ and $Z$) on the JT and $^{14}$N states by performing Ramsey experiments using different values of $\tau$ for preparation and measurement (Fig. \ref{fig3}b).

To investigate the spin-pair coherence, we measure the dephasing time for $\tau = 14.0$ \textmugreek s (Fig. \ref{fig3}c). We find $T_2^* = 44(9)$ ms, among the longest reported for solid-state electron-spin qubits \cite{Miao2020}. Compared to the single P1 electron-spin coherence $T_2^* = 50(3)$ \textmugreek s \cite{Degen2021}, this is a three-order-of-magnitude improvement in the same nuclear- and electron-spin bath. Two mechanisms contribute to this long dephasing time. First, the anti-parallel spin-pair states from a decoherence-free subspace: they are insensitive to the partially correlated noise from the P1 bath \cite{Bartling2022}. And second, the spin pair forms a clock transition due to the P1-P1 coupling \cite{Bartling2022} (Supplementary Sec. X).

\begin{figure}
\includegraphics[width = \columnwidth]{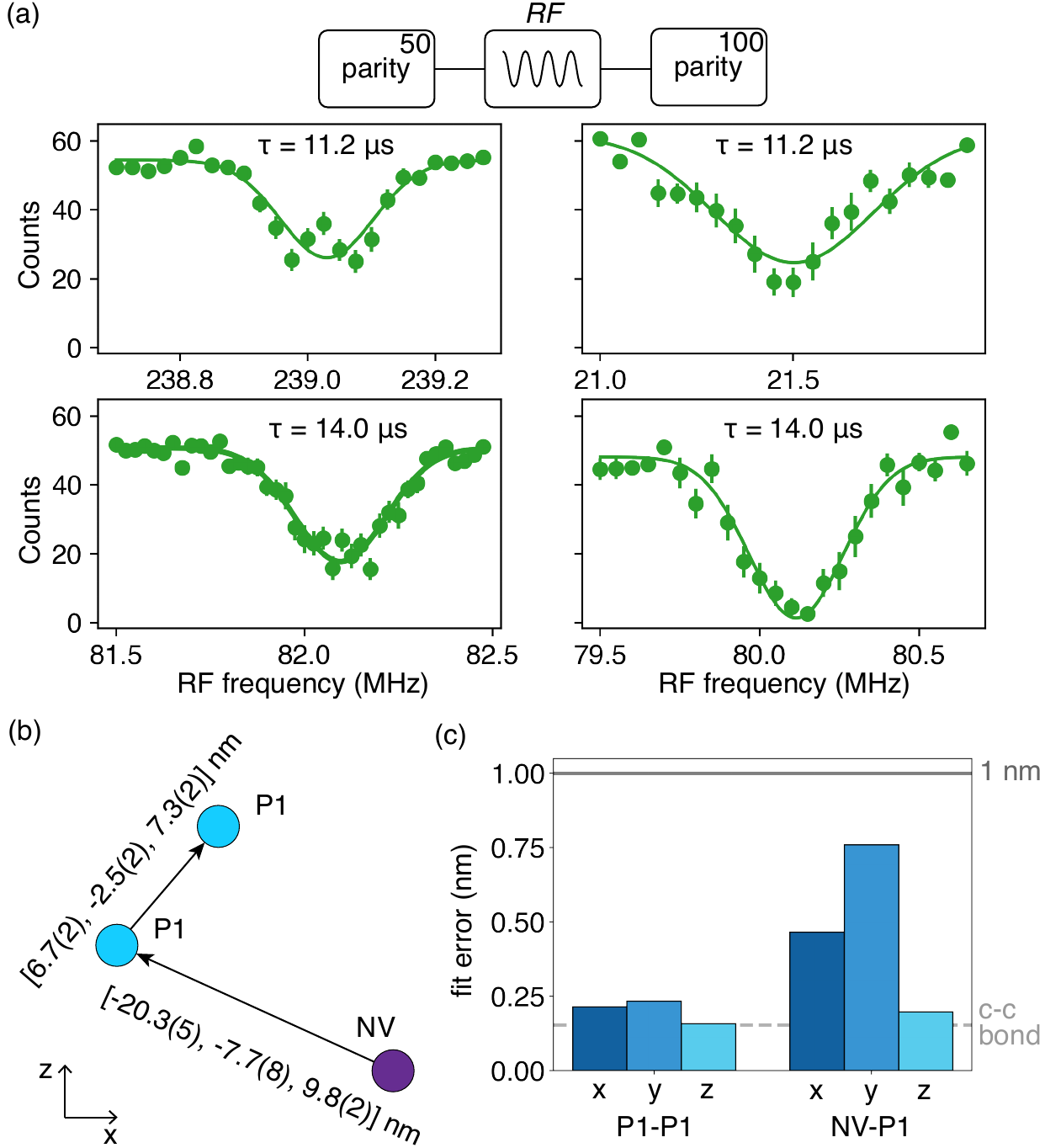}
\caption{\textbf{RF driving and imaging of the P1 electron-spin pair.} \textbf{(a)} (Top) Experimental sequence. The spin pair is initialized in $\frac{1}{2}\ket{\uparrow \downarrow}\bra{\uparrow \downarrow} + \frac{1}{2}\ket{\downarrow \uparrow} \bra{\downarrow \uparrow}$, after which an RF pulse of varying frequency can flip the spin pair to the parallel subspace (Supplementary Sec. IV). (Bottom) Spectra for $\tau = 11.2$ \textmugreek s (top) and $\tau = 14.0$ \textmugreek s (bottom). A reduction in counts gives information about the JT and $^{14}$N state probed for that $\tau$. \textbf{(b)} Fitted positions (up to inversion symmetry) of the two P1 centers (blue) with respect to the NV center (purple). \textbf{(c)} Fit errors for the P1-P1 and NV-P1 positions.}
\label{fig4}
\end{figure}

Next, we determine what JT and $^{14}$N states are associated to the signals for the different interpulse delays $\tau$. Due to the electron-nuclear hyperfine interaction and misaligned, finite magnetic field, the electron-spin transition frequency is different for each JT and $^{14}$N state. After initializing the electron-spin pair in $\frac{1}{2} \ket{\uparrow \downarrow}\bra{\uparrow \downarrow} + \frac{1}{2} \ket{\downarrow \uparrow} \bra{\downarrow \uparrow}$ for an unknown JT and $^{14}$N state, we apply a radio-frequency (RF) pulse that - when resonant - can flip the spin pair to the parallel subspace resulting in a change in signal on the NV center (Fig. \ref{fig4}a). The RF frequencies at which electron-spin flips occur, give information about the JT and $^{14}$N state associated to that $\tau$ (Supplementary Sec. IV).

Figure \ref{fig4}a shows the data for both $\tau = 11.2$ \textmugreek s and $\tau = 14.0$ \textmugreek s. In order to map the obtained frequencies to the JT and $^{14}$N state, we simulate the experiment for each JT and $^{14}$N configuration (Supplementary Sec. IV). From this, we obtain a set of possible Jahn-Teller axis and $^{14}$N spin state assignments (Supplementary Sec. IV). We perform the same analysis for $\tau = 16.8$ \textmugreek s, $\tau = 18.6$ \textmugreek s and $\tau = 29.0$ \textmugreek s (Supplementary Sec. V).

While the JT axis can be directly assigned, the $^{14}$N spin-state assignment is more complicated. Due to the electron-nitrogen spin mixing, flip-flops can occur that involve both the electron and nitrogen spin, creating additional possible transitions (Supplementary Sec. II). Next, we resolve this ambiguity by determining for which of the possible state assignments a consistent spatial structure of the system can be found.

The spin-pair couplings obtained in Fig. \ref{fig3}b combined with the associated JT and $^{14}$N states allow us to image the P1 electron-spin pair \footnote{As an alternative to using the internal degrees of freedom of the P1 center, one could sweep the external magnetic field angle \cite{Yudilevich2022, Cooper2020}.}. We fit the five measured P1-P1 couplings to different spatial configurations of the P1 pair. Multiple $^{14}$N spin-state assignments are possible for three out of five measured couplings (Supplementary Sec. V). We resolve this by considering all possible assignments in the fit. We benchmark the fitting algorithm on 10 randomly generated P1 pairs (Supplementary Sec. VII), after which we apply it to the measured couplings. The result yields the relative position of the two P1 centers with an uncertainty close to the diamond bond length (Fig. \ref{fig4}b,c), as well as the JT and $^{14}$N states associated to the $\tau$ resonances (Supplementary Sec. V) 

We find the position of the NV center, using a similar fitting algorithm (Supplementary Sec. VII). In Degen et al. \cite{Degen2021} double-resonance sequences were used to measure the dipolar couplings of the NV center to the two P1 centers. We fix the obtained P1-P1 relative position and fit to the NV-P1 couplings (up to inversion symmetry). The result is shown in Fig. \ref{fig4}b,c. 

A coherent interaction between the NV and the P1-pair requires the interaction time (set by $1/Z$) to be small compared to the spin-pair dephasing time ($T_2^*$) and the NV electron-spin coherence time under decoupling ($T_2$). A key element in our experiment is that the large number of different internal P1 states cause energy differences that prevent flip-flop dynamics in the bath. This extends the NV spin coherence and facilitates selective addressing of individual spin pairs, at the cost of a lower success probability for finding a given pair in the desired configuration. The full microscopic dynamics including the P1 and $^{13}$C nuclear-spin bath are complex. Predicting the typical number of P1 spin pairs that would be observed for randomly drawn NV centers (i.e. bath instances), likely requires detailed numerical simulation, which we do not pursue here. 

In conclusion, we experimentally demonstrated the detection, imaging and control of an electron-spin pair in a spin bath through the back-action from a central spin. These results experimentally access the underlying microscopic quantum dynamics which are central to theoretical methods, such as correlated cluster expansion (CCE), that have been widely used to understand time- and ensemble-averaged measurements \cite{Zhao2012Decoherence,Ye2019,Seo2016,Witzel2010,Witzel2012,Park2022}. The long dephasing times indicate electron-spin pairs based on P1 centers or other defects \cite{Shi2013b, Cooper2020,Yamamoto2013,Rosenfeld2018} might be interesting qubits. While the added complexity from the P1 internal states limits its use as a qubit, this could be partly overcome by applying a large, aligned magnetic field (Supplementary Sec. II). Lastly, the presented methods could contribute to efforts towards atomic-scale magnetic resonance imaging of complex spin samples outside of the diamond by directly detecting and imaging spin pairs \cite{Janitz2022,Vandestolpe2023}. \\ 

\begin{acknowledgments}
We thank V.V. Dobrovitski and M.H. Abobeih for discussions. This work was supported by the Netherlands Organisation for Scientific Research (NWO/OCW) through a Vidi grant, as part of the Frontiers of Nanoscience (NanoFront) programme and through the project QuTech Phase II funding: ``Quantum Technology for Computing and Communication” (Project No. 601.QT.001). This project has received funding from the European Research Council (ERC) under the European Union’s Horizon 2020 research and innovation programme (grant agreement No. 852410). This project (QIA) has received funding from the European Union’s Horizon 2020 research and innovation programme under grant agreement No. 820445. The fitting algorithms were performed on the DelftBlue supercomputer \cite{DHPC2022}.
\end{acknowledgments}

\bibliography{main.bib}

\end{document}


\begin{center}
    {\Large \centering \bf Supplemental Material}
\end{center}

\tableofcontents
\clearpage

\section{System Hamiltonian} \label{hamiltonians}
In this section, we discuss the Hamiltonian that describes the dynamics of the NV-P1-P1 system. Then, we explain how we calculate the effective coupling between the defects' electron spins. Lastly, we examine the possible orientations of the P1 centers in the diamond lattice.

\subsection{NV-P1-P1 Hamiltonian}
The Hamiltonian of the NV-P1-P1 system consists of the individual Hamiltonians of the NV and P1 centers and the Hamiltonians for the dipolar interactions between their respective electron spins \cite{Smith1959,Degen2021}. Note that we omit the Hamiltonian term that describes the NV nitrogen spin since the zero-field splitting suppresses spin mixing between the NV nitrogen and electron spin and the external magnetic field is (nearly) aligned along the NV symmetry axis. We also omit any Hamiltonian terms that describe the dipolar coupling between the $^{14}$N nuclear spin of one defect and the electron or $^{14}$N nuclear spin from other defects, due to the large difference in gyromagnetic ratios ($\gamma_e/\gamma_n\approx9000$) \cite{Degen2021}.
\begin{equation}
    H = H_{\text{NV}} + \sum^{2}_{j=1}H_{\text{P1,j}}+\sum^{3}_{j<k}H_{D_{jk}}
\end{equation}
where 
\begin{equation}
\begin{split}
    H_{\text{NV}} &= \Delta \hat{S}_z^2 + \gamma_e\mathbf{B}\cdot\mathbf{S}\\
    H_{\text{P1,j}}&=\gamma_e \mathbf{B}\cdot\mathbf{J}_j +  \gamma_n\mathbf{B}\cdot\mathbf{I}_j+
    \mathbf{J}_j\cdot {\mathbf{A}}_i\cdot\mathbf{I}_j +
    \mathbf{I}_j\cdot{\mathbf{P}}_i\cdot\mathbf{I}_j\\
    H_{D_{\text{P1,j}-\text{NV}}} &= D\cdot(3(\mathbf{J}_j\cdot\hat{r})(\mathbf{S}\cdot\hat{r}) - \mathbf{J}_j\cdot\mathbf{S})\\
    H_{D_{\text{P1,j}-\text{P1,k}}} &= D\cdot(3(\mathbf{J}_j\cdot\hat{r})(\mathbf{J}_k\cdot\hat{r}) - \mathbf{J}_j\cdot\mathbf{J}_k)\\
    \end{split}
\end{equation}

\begin{table*}[h]
\centering
\begin{tabular}{lll}
    $\mathbf{I}_j$ &=
    $\begin{bmatrix}
        \hat{I}_x, & \hat{I}_y, & \hat{I}_z 
    \end{bmatrix}_j$ 
    &\text{\hspace{1cm}Spin-1 operators for the $j$th P1 $^{14}$N nuclear spin}\\
     $\mathbf{J}_j$&= 
    $\begin{bmatrix}
        \hat{J}_x, & \hat{J}_y, & \hat{J}_z 
    \end{bmatrix}_j$ 
    &\text{\hspace{1cm}Spin-1/2 operators for the $j$th P1 electron spin}\\
    $\mathbf{S}$ &= 
    $\begin{bmatrix}
        \hat{S}_x, & \hat{S}_y, & \hat{S}_z
    \end{bmatrix}$
    &\text{\hspace{1cm}Spin-1 operators for the NV electron spin}\\
    $\mathbf{B}$ &=
    $\begin{bmatrix}
        B_x, & B_y, & B_z
    \end{bmatrix}$
    &\text{\hspace{1cm}B-field vector}\\
    ${\mathbf{A}}_i$ &= $ R^{T} A_{diag} R$  
    &\text{\hspace{1cm}Hyperfine coupling tensor} \\
    ${\mathbf{P}}_i$ &= $R^{T} P_{diag} R$ 
    &\text{\hspace{1cm}Quadrupolar coupling tensor}
\end{tabular}
\end{table*}

with $$D = \frac{-\mu_0 \gamma_e^2\hbar}{4\pi r^3}.$$ $\mu_0$ is the vacuum magnetic permeability, $\gamma_e\approx 2.8024$ MHz/G and $\gamma_n\approx 0.3078$ kHz/G are the gyromagnetic ratios of the electron and $^{14}$N spin respectively. $r$ is the physical separation of the electron spins and $\hat{r}=\mathbf{r}/\abs{r}$ the unit vector between them. $R$ are the rotation matrices of the $SO(3)$ group, with Euler angles $\alpha, \beta, \gamma$ \cite{Degen2021,Smith1959}:
\begin{equation}
    \footnotesize
    R(\alpha,\beta,\gamma) = 
    \begin{pmatrix}
    \cos(\gamma)\cos(\beta)\cos(\alpha)-\sin(\gamma)\sin(\alpha) & \cos(\gamma)\cos(\beta)\sin(\alpha)+\sin(\gamma)\cos(\alpha) & -\cos(\gamma)\sin(\beta)\\
    -\sin(\gamma)\cos(\beta)\cos(\alpha)-\cos(\gamma)\sin(\alpha) & -\sin(\gamma)\cos(\beta)\sin(\alpha)+\cos(\gamma)\cos(\alpha) & \sin(\gamma)\sin(\beta) \\
    \sin(\beta)\sin(\alpha) & \sin(\beta)\sin(\alpha) & \cos(\beta)
    \end{pmatrix}
\end{equation}
The rotation matrices $R$ rotate the hyperfine and quadrupolar tensors depending on the Jahn-Teller axis, i.e ${\mathbf{A}}_{i} = R(\alpha,\beta)^{T} A_{diag} R(\alpha,\beta)$. The Euler angles $\alpha,\beta$ define the Jahn-Teller principal axis (Fig. \ref{fig:configurations}). Due to the axial symmetry of the P1 center in its principal axis ($A_x = A_y$ and $P_x = P_y$), we can set $\gamma=0$ without loss of generality \cite{Degen2021}. The rotation matrix $R$ simplifies to
\begin{equation}
    R(\alpha,\beta) = 
    \begin{pmatrix}
    \cos(\beta)\cos(\alpha) & \cos(\beta)\sin(\alpha) & -\sin(\beta) \\
    -\sin(\alpha) & \cos(\alpha) & 0 \\
    \sin(\beta)\cos(\alpha) & \sin(\beta)\sin(\alpha) & \cos(\beta)
    \end{pmatrix}
\end{equation}

The hyperfine and quadrupolar tensors are $A_{diag} = \text{diag}[114.03, 81.31, 81.31]$ MHz and $P_{diag} = \text{diag}[2.65, 1.32, 1.32 ]$ MHz \cite{Knowles2014}. Note that the effect of the NV-P1 dipolar coupling can be approximated as pure dephasing of the form $\hat{S}_z \hat{J}_z$ due to the large NV-P1 energy difference and the large NV and P1 Zeeman energies compared to the NV-P1 coupling. 

\subsection{Effective coupling}
Now, we describe how we calculate the effective couplings between the different defect spins (P1-P1 and NV-P1). These effective couplings provide the parameters $X_{(i, m_I)}$ and $Z_{(i, m_I)}$ of the pseudo-spin Hamiltonian.

We start by constructing the Hamiltonian of the system of interest, i.e $H_{\text{NV-P1}}$ or $H_{\text{P1-P1}}$. To calculate the effective coupling between the electron spins, we consider the energy differences between their eigenstates. For example, for the NV-P1 system, with the P1 center in a particular Jahn-Teller axis $i$ and its $^{14}$N nuclear spin in $\ket{+}$ ($m_I = +1$), the effective coupling between the electron spins is given by \cite{Degen2021}
\begin{equation}
    D_{+,i}=(\lambda_{\ket{-1,+,\downarrow},i}-\lambda_{\ket{-1,+,\uparrow},i})+ (\lambda_{\ket{0,+,\uparrow},i}-\lambda_{\ket{0,+,\downarrow},i}),
\end{equation}
where $\lambda$ are the eigenvalues corresponding to the eigenvectors denoted in their subscripts. The indicated eigenstates correspond to the NV electron spin, the P1 nuclear spin, and the P1 electron spin respectively.

For two P1 centers undergoing flip-flop dynamics, the effective coupling is given by
\begin{equation}
    X=\lambda_{\frac{1}{\sqrt{2}}(\ket{+,\downarrow,+,\uparrow}+\ket{+,\uparrow,+,\downarrow})}-\lambda_{\frac{1}{\sqrt{2}}(\ket{+,\downarrow,+,\uparrow}-\ket{+,\uparrow,+,\downarrow})},
\end{equation}
where for this example the nitrogen spin for both P1 centers is in the $\ket{+}$ ($m_I = +1$) state. The indicated eigenstates correspond to the nitrogen and electron spin of the first P1 center respectively, followed by the nitrogen and electron spin of the second P1 center.

\subsection{P1 center orientations}
\label{subsec:orientation}
\begin{figure}[b]
\begin{minipage}[c]{0.4\textwidth}
\centering
\includegraphics[width=\textwidth]{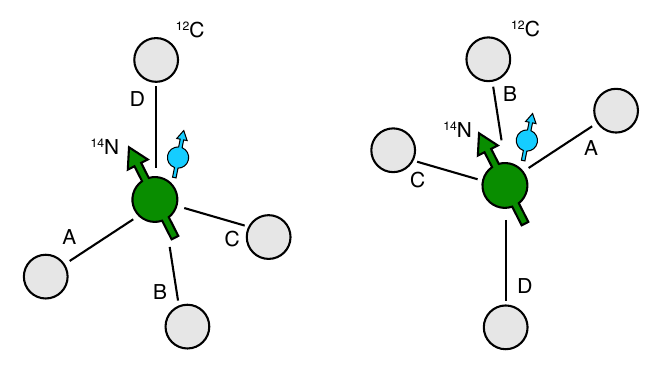}
\end{minipage}
\begin{minipage}[c]{0.45\textwidth}
\begin{tabular}[b]{c|cc|cc}
\centering
JT & $\alpha$ & $\beta$ & $\alpha'$ & $\beta'$\\
\hline
A & 0 & 109.5 & 180 & 70.5\\
B & 120 & 109.5 & 300 & 70.5 \\
C & 240 & 109.5 & 60 & 70.5 \\
D & 0 & 0 & 180 & 180 \\
\end{tabular}
\end{minipage}
\caption{\textbf{Schematic of the two possible orientations of a P1 center with corresponding rotation angles.} On the left, the two possible orientations of a P1 center in the diamond lattice are shown. The Jahn-Teller axes are indicated by $A, B, C$ and $D$. The Euler angles corresponding to these Jahn-Teller axes are given in degrees on the right, where the left (right) column corresponds to the left (right) orientation.}
\label{fig:configurations}
\end{figure}

P1 centers appear in two different orientations in the diamond lattice, as depicted in Fig. \ref{fig:configurations}. In this section, we show that these two different orientations do not result in an observable difference in our experiments. The hyperfine and quadrupolar tensors are transformed differently by the two Euler angles $\alpha$ and $\beta$ depending on the Jahn-Teller axis. In the table in Fig. \ref{fig:configurations}, the first (second) set of $\alpha$, $\beta$ ($\alpha'$, $\beta'$) corresponds to the left (right) picture of the P1 center orientations.

We now show that the transformation of a diagonal matrix $M_i = R^{T}(\alpha_i,\beta_i) M_{diag} R(\alpha_i,\beta_i)$ is equivalent for both orientations $(\alpha_i,\beta_i)$ and $(\alpha_i',\beta_i')$. Let $R=R(\alpha,\beta)$ and $R'=R(\alpha',\beta')$. We can then write
\begin{equation}
\begin{split}
    R'M_{i}R'^T&=R'R^{T} M_{diag} R R'^T\\ 
    &=M_{diag} R'R^{T} R R'^T\\
    &= M_{diag}   \\
    \implies M_{i} &= R'^{T} M_{diag} R'\\
    \end{split}
\end{equation}
Here we made use of the fact that $R'R^{T}=\text{diag}(1,-1,-1)$ is a diagonal matrix for $\alpha' = \alpha + \pi$ and $\beta' = \pi - \beta$. It thus commutes with the diagonal matrix $M_{diag}$. Since the dipole term in the Hamiltonian is also invariant under the different Jahn-Teller distortions, the couplings describing the dynamics remain unchanged for the two orientations, and we do not expect an observable difference in our measurements. 

\section{Effects of P1 electron-nitrogen spin mixing} \label{mixing}

In the following, we will denote the single P1 eigenbasis as $\ket{- \downarrow}$, $\ket{- \uparrow}$, $\ket{0 \downarrow}$, $\ket{0 \uparrow}$, $\ket{+ \downarrow}$, $\ket{+ \uparrow}$, where the first entry refers to the nitrogen spin ($m_I = -1, 0, +1$) and the second entry refers to the electron spin ($\uparrow$ or $\downarrow$). In a simple picture, one would expect to observe flip-flop dynamics of two spins if they are degenerate and have nonzero dipolar coupling. For example, for a pair of P1 centers the states $\ket{- \downarrow - \uparrow}$ and $\ket{- \uparrow - \downarrow}$ are degenerate. The dipolar coupling between the two electron spins can induce flip-flops between them. Then, the eigenstates become $\frac{1}{\sqrt{2}} (\ket{- \downarrow - \uparrow} \pm \ket{- \uparrow - \downarrow})$. Consider two other states of the pair of P1 centers: $\ket{- \downarrow 0 \uparrow}$ and $\ket{0 \uparrow - \downarrow}$. These two states are degenerate as well. However, in principle, we do not get any flip-flop dynamics, because the dipolar coupling between the two electron spins cannot induce nitrogen spin flips.

This simple picture holds when the magnetic field is much larger than the hyperfine interaction with the P1 nitrogen spin-1 ($\gamma_e B \gg A_\parallel, A_\perp$). Due to the finite magnetic field we work at ($\gamma_e B \sim A_\parallel, A_\perp$), the single P1 eigenstates are not separable into a nitrogen spin-1 and an electron spin-1/2 part. Instead, the nitrogen and electron spin become mixed. The finite magnetic field mostly leads to mixing of \{$\ket{- \uparrow}$, $\ket{0 \downarrow}$\} and \{$\ket{0 \uparrow}$, $\ket{+ \downarrow}$\}. This implies that the dipolar coupling between the two electron spins can induce flip-flop dynamics between states such as $\ket{0 \downarrow - \uparrow}$ and $\ket{- \uparrow 0 \downarrow}$.

Therefore, we expect to observe signal on the NV center due to two types of flip-flop states. First, there is the simple case in which the nitrogen spin states are approximately equal and fixed, and the electron spins form superpositions of $\ket{\uparrow \downarrow}$ and $\ket{\downarrow \uparrow}$. Second, the mixing between the nitrogen spin and electron spin of the P1 centers allows for more complicated flip-flop dynamics.

\begin{figure}
\includegraphics[width = 0.8\columnwidth]{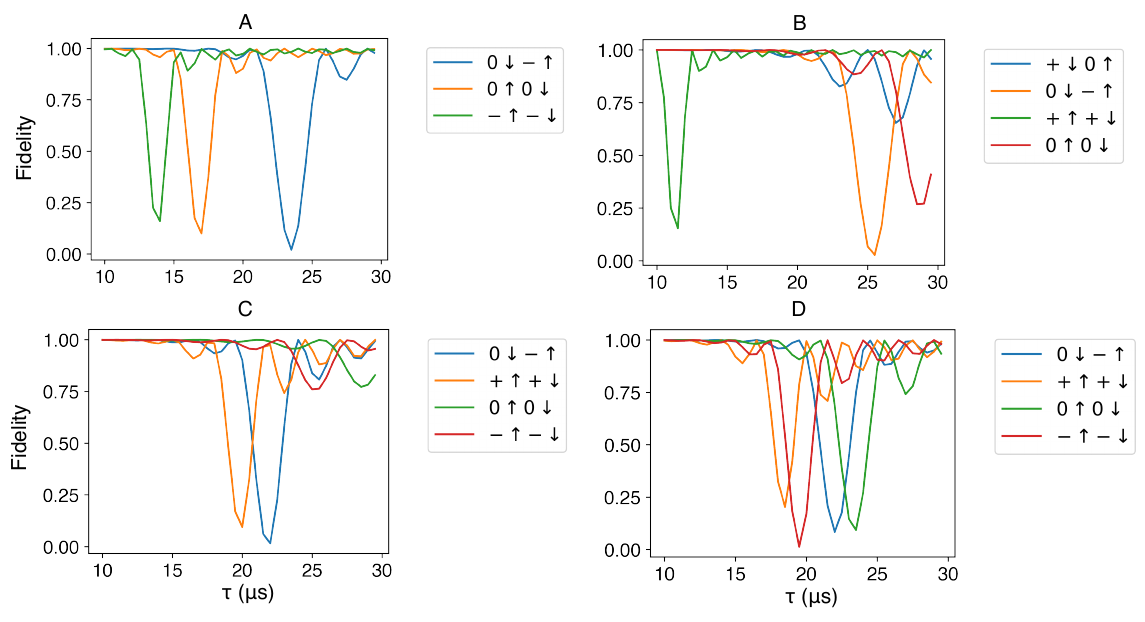}
\caption{\textbf{Simulation of the effect of a pair of P1 centers on dynamical decoupling of the NV electron spin grouped per Jahn-Teller state.} We simulate the system discussed in this paper: a single NV center coupled to two P1 centers. We use the same magnetic field as in the experiments: $\mathbf{B} = [2.43,1.42,45.552]$ G. The position vectors for the NV center and two P1 centers are the ones extracted in the main text. Then, we consider the two P1 centers to be in the same Jahn-Teller state ($A$, $B$, $C$ or $D$) and calculate the expected signal on the NV center for each P1-P1 eigenvector and for each Jahn-Teller state. Each of the four figures corresponds to a different Jahn-Teller state, indicated by the title. We only show states that lead to a significant coherence loss on the NV electron spin: less than a fidelity of 0.9. The legends on the right of each graph show which P1-P1 states cause the NV electron coherence loss. We observe coherence loss from both types of flip-flop states: when the nitrogen is fixed as well as more complicated flip-flop states that involve nitrogen spin flips. The latter states can give significant signal, comparable to the flip-flop states that do not involve the nitrogen spin.}
\label{DD_single_JT_sim}
\end{figure}

In Fig. \ref{DD_single_JT_sim} we show the simulated dynamical decoupling signal on the NV center per Jahn-Teller state for the NV-P1-P1 positions obtained in this work. For each Jahn-Teller state, we calculate the expected NV electron signal per P1-P1 eigenvector. Only eigenvectors generating significant signal (fidelity loss of $ > 0.1$) are shown. The labels are obtained by calculating the overlap of the P1-P1 eigenvector with the spin basis vectors and indicating the largest overlap.

From Fig. \ref{DD_single_JT_sim} it can be seen that most NV electron coherence loss due to P1 electron-spin pairs is due to the simple flip-flop states: $\ket{+ \uparrow + \downarrow}, \ket{0 \uparrow 0 \downarrow}, \ket{- \uparrow - \downarrow}$. However, there are more complicated flip-flop states due to the mixing between the nitrogen spin and electron spin of the P1 centers, amongst which are $\ket{0 \downarrow - \uparrow}$ and $\ket{+ \downarrow 0 \uparrow}$.

\begin{figure}
\includegraphics[width = 0.8\columnwidth]{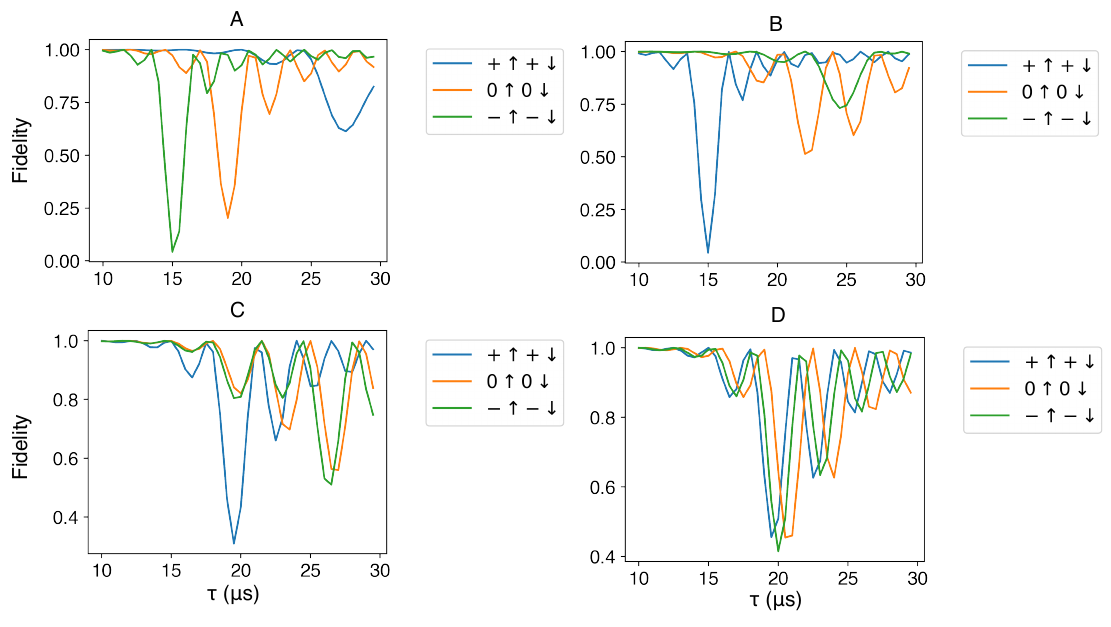}
\caption{\textbf{Simulation of the effect of a pair of P1 centers on dynamical decoupling of the NV electron spin grouped per Jahn-Teller state at a high magnetic field.} We simulate the same system as in Fig. \ref{DD_single_JT_sim} at a magnetic field of $[B_x, B_y, B_z] = [1,1,100]$ G. At magnetic fields significantly greater than the P1 electron-nitrogen hyperfine coupling, only simple flip-flop states generate signal on the NV center.}
\label{DD_single_JT_sim_highB}
\end{figure}

To further illustrate the origin of the more complicated flip-flop states, we simulate the same system at a high magnetic field ($\mathbf{B} = [1,1,100]$ G). It is expected that only the simple flip-flop states remain, because in this regime it holds that $\gamma_e B > A_\parallel, A_\perp$, resulting in negligible P1 electron-nitrogen spin mixing. The result of the simulation is shown in Fig. \ref{DD_single_JT_sim_highB}, where we do indeed find that only the simple flip-flop states remain.

At the magnetic field used in the experiments in this paper ($B \sim 45$ G), the P1 electron-nitrogen spin mixing is not negligible. Hence, we have to consider the possibility of measuring dipole-dipole couplings resulting from electron-nitrogen spin mixing. To address this possibility, we also fit the measurements to these flip-flop states, see Supplementary Section \ref{imaging}. 

\begin{figure}
\includegraphics[width = 0.8\columnwidth]{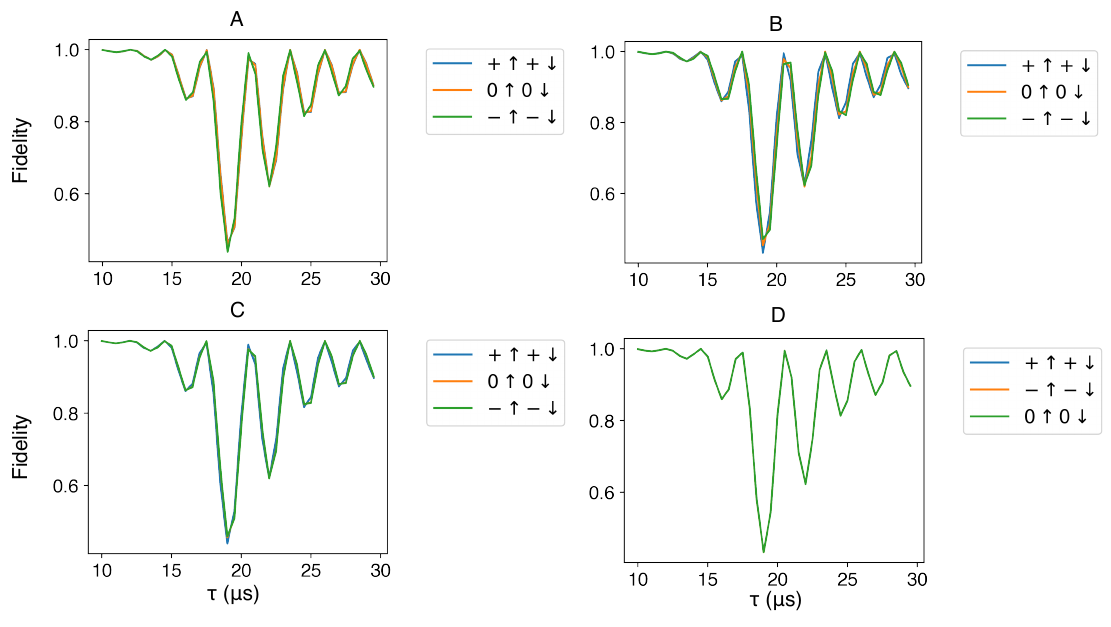}
\caption{\textbf{Simulation of the effect of a pair of P1 centers on dynamical decoupling of the NV electron spin grouped per Jahn-Teller state at a very high magnetic field.} We simulate the same system as in Fig. \ref{DD_single_JT_sim} at a magnetic field of $[B_x, B_y, B_z] = [100,100,10000]$ G. At this magnetic field, the external magnetic field is significantly greater than the electron-nuclear hyperfine interaction and the nuclear quadrupole interaction. Hence, we observe close to the same signal for all Jahn-Teller axes and all nitrogen-spin states.}
\label{DD_single_JT_sim_very_highB}
\end{figure}

For completeness, we simulate the same system at a very high, but misaligned magnetic field of $\mathbf{B} = [100,100,10000]$ G. The result is shown in Fig. \ref{DD_single_JT_sim_very_highB}. We observe that all Jahn-Teller axes and nitrogen-spin states give approximately the same signal. For these values of the magnetic field, the effect of the electron-nitrogen spin mixing is close to negligible and the P1-P1 electron-electron coupling is therefore almost the same for all configurations. 

Importantly, the two P1 centers still need to be in the same Jahn-Teller axis and nitrogen spin state to be degenerate and to flip-flop. However, the amplitude of a dip in the dynamical decoupling signal goes up from 1/288 to 1/24. The ratio of 1/288 comes from 4 Jahn-Teller axes for each P1 center, 3 nitrogen spin states for each P1 center and 2 electron spin states for each P1 center, two of which ($\ket{\uparrow \downarrow}, \ket{\downarrow \uparrow}$) generate signal. Since all 12 possibilities in Fig. \ref{DD_single_JT_sim_very_highB} give the same signal, the amplitude goes up from 1/288 to 1/24.

When the external magnetic field is very high and aligned, the Jahn-Teller axes $A$, $B$ and $C$ also become degenerate. This increases the fraction $1/24$ further to $5/48$, since six additional Jahn-Teller configurations exhibit flip-flop dynamics at the flip-flop rate.

\clearpage

\section{Spin and parity readout} \label{spin_parity_readout}

\begin{figure}[h]
\includegraphics[width = \columnwidth]{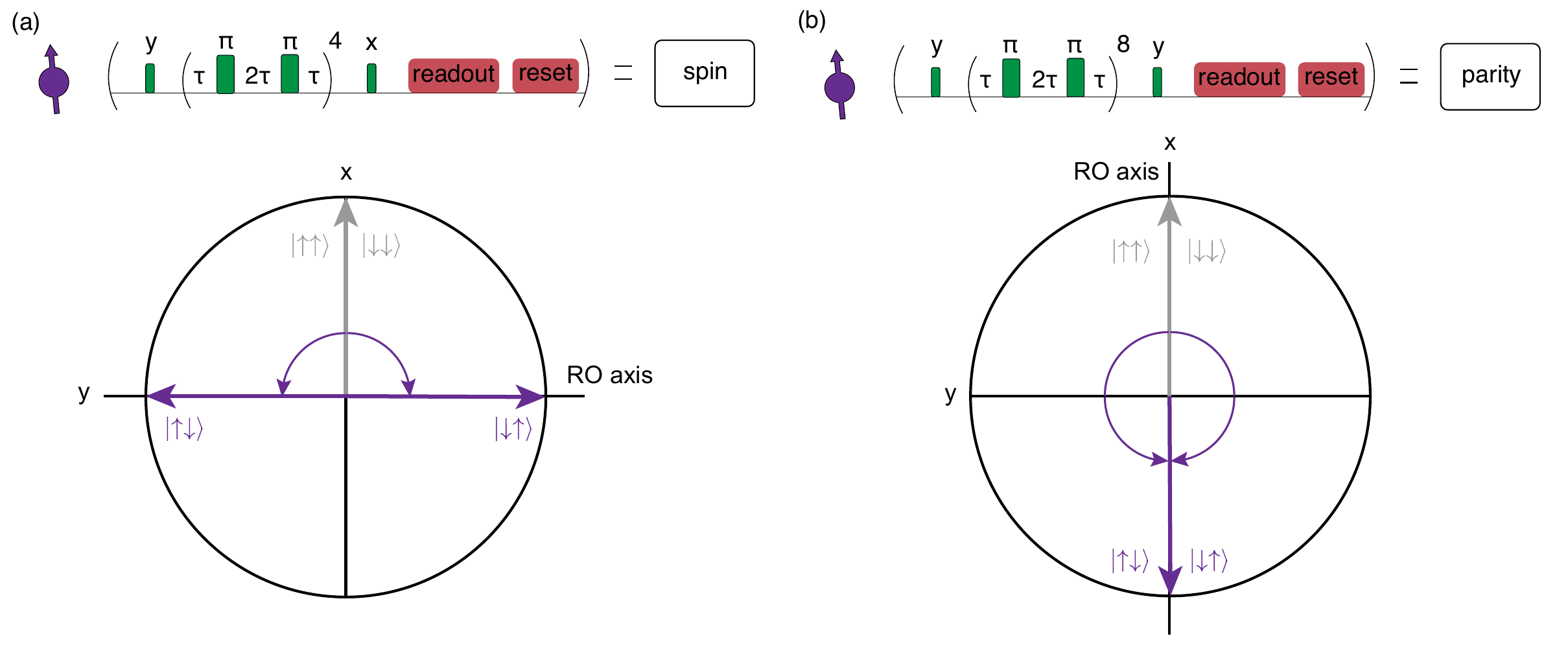}
\caption{\textbf{Evolution of the NV center during spin and parity readout.} \textbf{(a)} Spin readout sequence and the corresponding phase pick-up of the NV center for the four different spin-pair states shown on a 2D Bloch sphere. We calibrate the number of dynamical decoupling units such that the NV electron spin picks up a $\pm \pi/2$ phase for $\ket{\uparrow \downarrow}$ and $\ket{\downarrow \uparrow}$. Then, we read out along the $y$-axis. \textbf{(b)} Parity readout sequence and the corresponding phase pick-up of the NV center for the four different spin-pair states shown on a 2D Bloch sphere. We calibrate the number of dynamical decoupling units such that the NV electron spin picks up a $\pm \pi$ phase for $\ket{\uparrow \downarrow}$ and $\ket{\downarrow \uparrow}$. Then, we read out along the $x$-axis.}
\label{bloch_spheres}
\end{figure}

In Fig. \ref{bloch_spheres}, we show the spin and parity readout together with Bloch spheres indicating the phase picked up by the NV center electron spin. When the P1 electron-spin pair is in the parallel state ($\{\ket{\uparrow \uparrow}, \ket{\downarrow \downarrow}\}$), there are no flip-flop dynamics and the NV electron spin does not pick up any phase. However, when the P1 electron-spin pair is in the anti-parallel subspace (\{$\ket{\uparrow \downarrow}$, $\ket{\downarrow \uparrow}$\}), the NV picks up a positive or negative phase depending on the pseudo-spin state. Note that the NV electron spin only picks up phase when the interpulse delay $\tau$ is resonant with the P1 pair flip-flop dynamics (and when the P1 pair is thus in that particular JT and $^{14}$N configuration).

For the spin readout, we tune the number of dynamical decoupling units such that the NV electron spin picks up a phase of $\pm \pi/2$ for the two anti-parallel spin-pair states $\ket{\uparrow \downarrow}$ and $\ket{\downarrow \uparrow}$. If we then read out along the $y$-axis, we can distinguish between $\ket{\uparrow \downarrow}$ and $\ket{\downarrow \uparrow}$. For the parity readout, we use double the number of dynamical decoupling units such that the NV electron spin picks up a phase of $\pm \pi$. If we read out along the $x$-axis, we can distinguish between the spin pair being in the parallel and anti-parallel subspace. By combining parity and spin readouts (Fig. 3), we can initialize the P1 spin pair in a specific anti-parallel state.
In the final readout, we use spin readouts to distinguish between the two anti-parallel states (Fig. 3).

\clearpage 

\section{RF simulations} \label{rf_simulations}

In this section, we simulate the application of a radio-frequency (RF) pulse on a single P1 center. To take into account the electron-nitrogen interaction in a single P1 center, we simulate the full time-dependent Hamiltonian under application of a single RF pulse by adding the Hamiltonian term

\begin{equation}
    H_{\text{RF}} = \Omega \cos (2 \pi f t + \phi) \hat{J}_x + \frac{\gamma_n}{\gamma_e} \Omega \cos (2 \pi f t + \phi) \hat{I}_x
\end{equation}

where $\gamma_e$ ($\gamma_n$) is the electron (nitrogen) spin gyromagnetic ratio. For the simulations, we set $\Omega = 250$ kHz, comparable to the Rabi frequency in the experiment (Supplementary Sec. \ref{rabi}). Then, $\Omega \gg X$ which means we can neglect the effect of the P1-P1 dipole-dipole coupling under the application of an RF pulse. Thus, it suffices to simulate the application of an RF pulse on a single P1 center, which makes the simulation of the time-dependent Hamiltonian significantly faster.

\begin{figure*}
\includegraphics[width = \columnwidth]{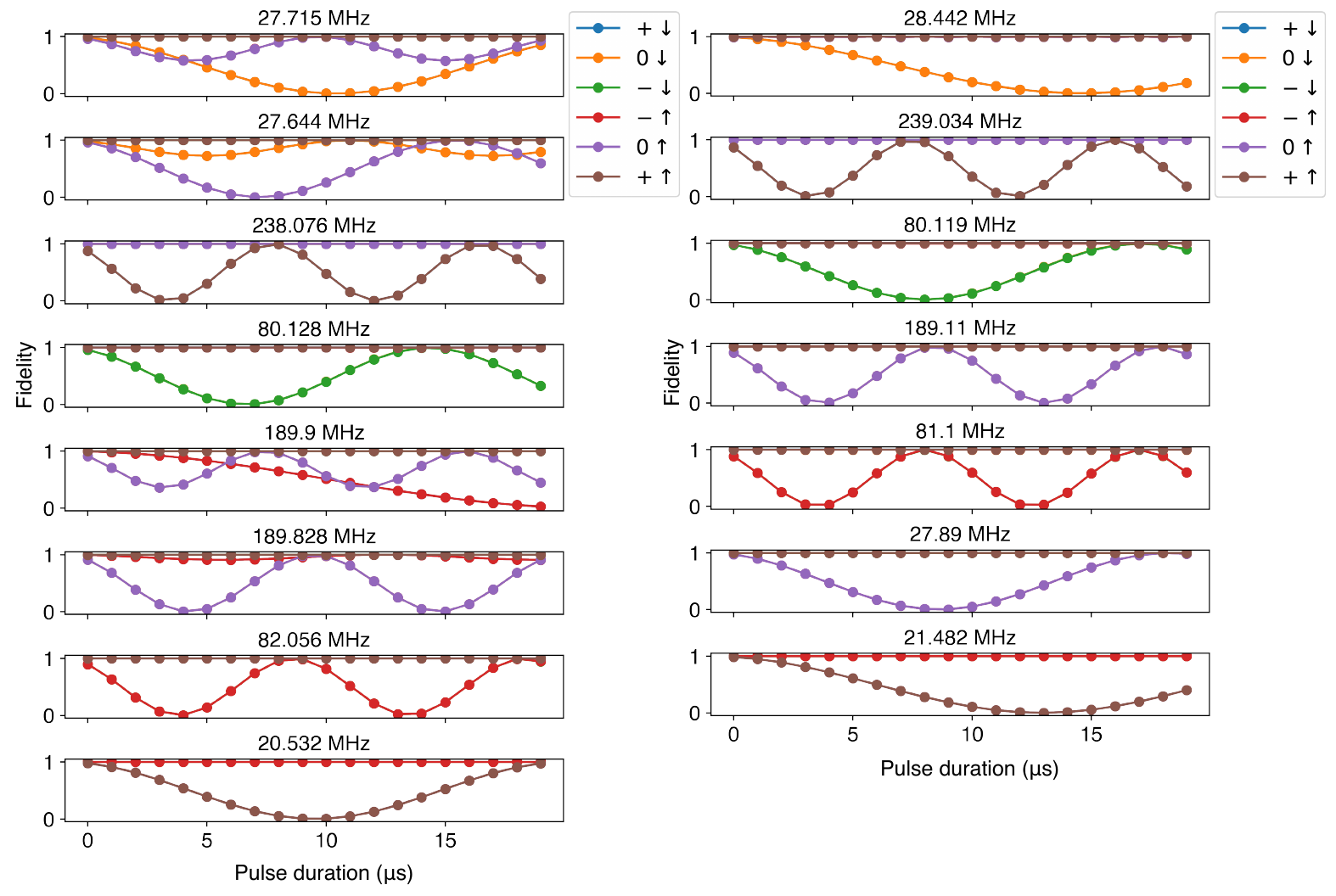}
\caption{\textbf{Simulation of Rabi oscillations of a P1 center for various RF frequencies and initial states.} Rabi oscillations are simulated for two different Jahn-Teller axes $A$ (left) and $B$ (right). The frequency at which the RF pulse is applied is indicated on top of each plot. For each plot, we simulate the application of the RF pulse for each eigenstate. We denote the eigenstates with $-/0/+$ indicating the (approximate) nitrogen-spin state and $\uparrow/\downarrow$ indicating the (approximate) electron-spin state. The signals originating from particular eigenstates can be the same, which makes their Rabi oscillations overlap. We clarify this by converting the simulated Rabi oscillations to Tables \ref{truth_table_A}, \ref{truth_table_B}.}
\label{rf_rabi_simAB}
\end{figure*}

\begin{figure*}
\includegraphics[width = \columnwidth]{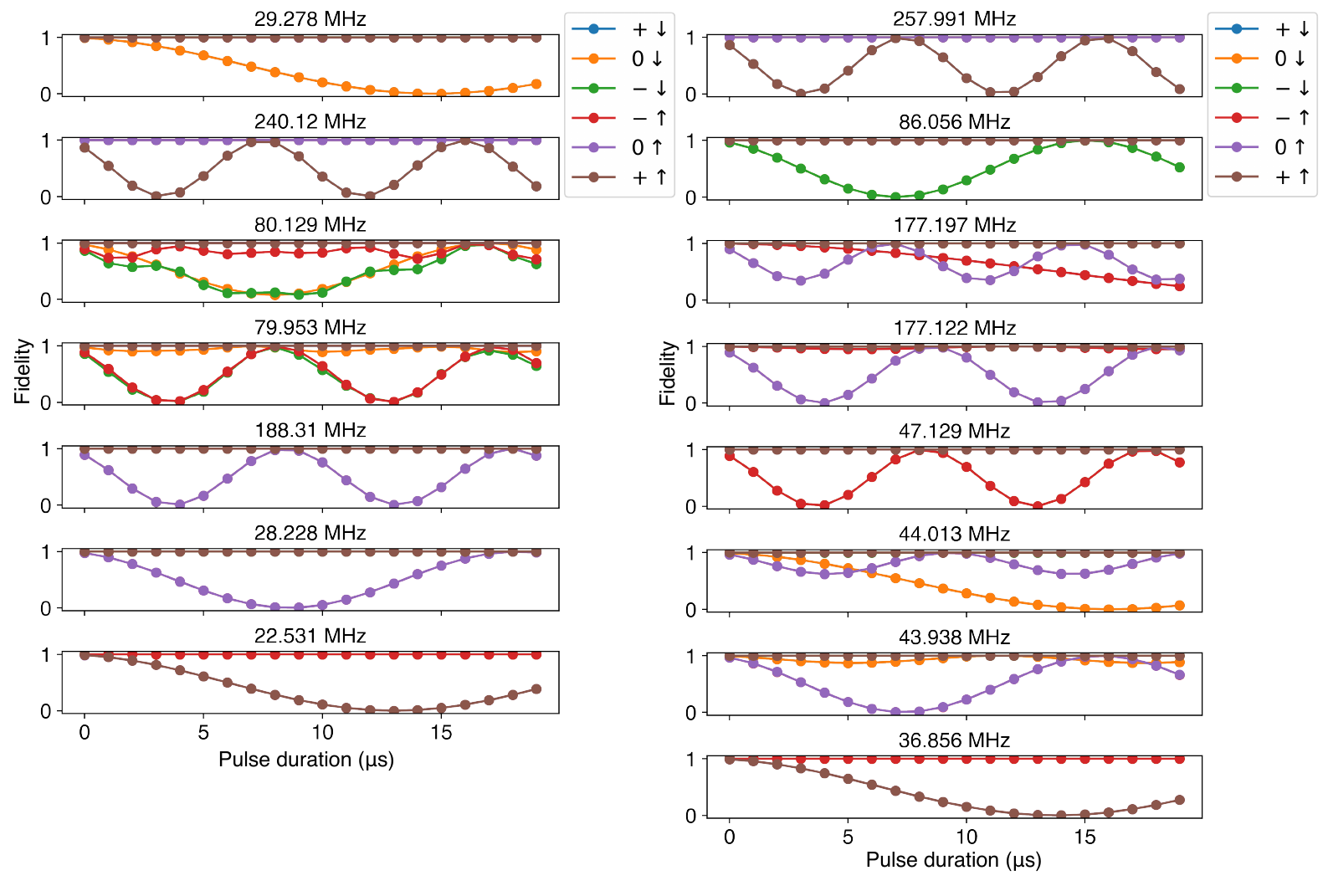}
\caption{\textbf{Simulation of Rabi oscillations of a P1 center for various RF frequencies and initial states.} Rabi oscillations are simulated for two different Jahn-Teller axes $C$ (left) and $D$ (right). The frequency at which the RF pulse is applied is indicated on top of each plot. For each plot, we simulate the application of the RF pulse for each eigenstate. We denote the eigenstates with $-/0/+$ indicating the (approximate) nitrogen-spin state and $\uparrow/\downarrow$ indicating the (approximate) electron-spin state. The signals originating from particular eigenstates can be the same, which makes their Rabi oscillations overlap. We clarify this by converting the simulated Rabi oscillations to Tables \ref{truth_table_C}, \ref{truth_table_D}.}
\label{rf_rabi_simCD}
\end{figure*}

In Figures \ref{rf_rabi_simAB}, \ref{rf_rabi_simCD} the results are shown for each of the four Jahn-Teller axes separately. For each relevant RF frequency, we simulate the evolution of the system for the six different eigenstates the P1 center can be in. For a particular frequency, certain eigenstates will give signal but others do not. Hence, observing a Rabi oscillation gives information about the Jahn-Teller axis as well as the nitrogen-spin state of the P1 center.

Next, we convert Figs. \ref{rf_rabi_simAB}, \ref{rf_rabi_simCD} to truth tables in order to make it straightforward to compare against experiment. If the Rabi oscillation of a particular eigenstate dips below 0.95, we consider that to be an observable signal and indicate it with a ``1'' in Tables \ref{truth_table_A}, \ref{truth_table_B}, \ref{truth_table_C}, \ref{truth_table_D}. If the Rabi oscillation does not dip below 0.95, we do not consider that to be an observable signal and indicate it with a ``0'' in Tables \ref{truth_table_A}, \ref{truth_table_B}, \ref{truth_table_C}, \ref{truth_table_D}. The frequencies between the different tables are different, since each Jahn-Teller axis has different eigenfrequencies due to the misaligned magnetic field. This makes it relatively straightforward to determine the Jahn-Teller axis. To determine the combination of eigenstates that cause flip-flop dynamics, we use a combination of the measurements and the fitting procedure (Supplementary Sec. \ref{imaging}).

\clearpage

\begin{table}
\small
\centering
\begin{tabular}{ |c|c|c|c|c|c|c| } 
 \hline
 $f$ (MHz) & $\ket{+ \downarrow}$ & $\ket{0 \downarrow}$ & $\ket{- \downarrow}$ & $\ket{- \uparrow}$ & $\ket{0 \uparrow}$ & $\ket{+ \uparrow}$ \\  \hline
 27.645/27.715 & 1 & 1 & 0 & 1 & 1 & 0 \\  \hline
 238.079 & 1 & 0 & 0 & 0 & 0 & 1  \\ \hline
 80.127 & 0 & 1 & 1 & 0 & 0 & 0  \\ \hline
 189.902 & 1 & 1 & 0 & 1 & 1 & 0  \\ \hline
 189.831 & 1 & 1 & 0 & 1 & 1 & 0  \\ \hline
 82.06 & 0 & 0 & 1 & 1 & 0 & 0  \\ \hline
 20.532 & 0 & 0 & 0 & 0 & 1 & 1  \\ \hline
\end{tabular}
\caption{\textbf{Truth table for Jahn-Teller state A.} For each frequency, it is indicated per (approximate) P1 eigenstate whether a Rabi oscillation is observed (``1'') or not (``0'').}
\label{truth_table_A}
\end{table}

\begin{table}
\small
\centering
\begin{tabular}{ |c|c|c|c|c|c|c| } 
 \hline
 $f$ (MHz) & $\ket{+ \downarrow}$ & $\ket{0 \downarrow}$ & $\ket{- \downarrow}$ & $\ket{- \uparrow}$ & $\ket{0 \uparrow}$ & $\ket{+ \uparrow}$ \\  \hline
 28.441 & 1 & 1 & 0 & 0 & 0 & 0 \\  \hline
 239.035 & 1 & 0 & 0 & 0 & 0 & 1  \\ \hline
 80.119 & 0 & 1 & 1 & 0 & 0 & 0  \\ \hline
 189.114 & 0 & 1 & 0 & 0 & 1 & 0  \\ \hline
 81.106 & 0 & 0 & 1 & 1 & 0 & 0  \\ \hline
 27.89 & 0 & 0 & 0 & 1 & 1 & 0  \\ \hline
 21.48 & 0 & 0 & 0 & 0 & 1 & 1  \\ \hline
\end{tabular}
\caption{\textbf{Truth table for Jahn-Teller state B.} For each frequency, it is indicated per (approximate) P1 eigenstate whether a Rabi oscillation is observed (``1'') or not (``0'').}
\label{truth_table_B}
\end{table}

\begin{table}
\small
\centering
\begin{tabular}{ |c|c|c|c|c|c|c| } 
 \hline
 $f$ (MHz) & $\ket{+ \downarrow}$ & $\ket{0 \downarrow}$ & $\ket{- \downarrow}$ & $\ket{- \uparrow}$ & $\ket{0 \uparrow}$ & $\ket{+ \uparrow}$ \\  \hline
 29.281 & 1 & 1 & 0 & 0 & 0 & 0 \\  \hline
 240.127 & 1 & 0 & 0 & 0 & 0 & 1  \\ \hline
 80.128 & 0 & 1 & 1 & 1 & 0 & 0  \\ \hline
 79.952 & 0 & 1 & 1 & 1 & 0 & 0  \\ \hline
 188.31 & 0 & 1 & 0 & 0 & 1 & 0  \\ \hline
 28.23 & 0 & 0 & 0 & 1 & 1 & 0  \\ \hline
 22.535 & 0 & 0 & 0 & 0 & 1 & 1  \\ \hline
\end{tabular}
\caption{\textbf{Truth table for Jahn-Teller state C.} For each frequency, it is indicated per (approximate) P1 eigenstate whether a Rabi oscillation is observed (``1'') or not (``0'').}
\label{truth_table_C}
\end{table}

\begin{table}
\small
\centering
\begin{tabular}{ |c|c|c|c|c|c|c| } 
 \hline
 $f$ (MHz) & $\ket{+ \downarrow}$ & $\ket{0 \downarrow}$ & $\ket{- \downarrow}$ & $\ket{- \uparrow}$ & $\ket{0 \uparrow}$ & $\ket{+ \uparrow}$ \\  \hline
 257.994 & 1 & 0 & 0 & 0 & 0 & 1  \\ \hline
 86.055 & 0 & 1 & 1 & 0 & 0 & 0  \\ \hline
 177.2 & 1 & 1 & 0 & 1 & 1 & 0  \\ \hline
 177.125 & 0 & 1 & 0 & 0 & 1 & 0  \\ \hline
 47.132 & 0 & 0 & 1 & 1 & 0 & 0  \\ \hline
 43.938/44.013 & 1 & 1 & 0 & 1 & 1 & 0 \\  \hline
 36.856 & 0 & 0 & 0 & 0 & 1 & 1  \\ \hline
\end{tabular}
\caption{\textbf{Truth table for Jahn-Teller state D.} For each frequency, it is indicated per (approximate) P1 eigenstate whether a Rabi oscillation is observed (``1'') or not (``0'').}
\label{truth_table_D}
\end{table}

\clearpage

\section{RF Rabi oscillations} \label{rabi}

\begin{figure}
\includegraphics[width = 0.9\columnwidth]{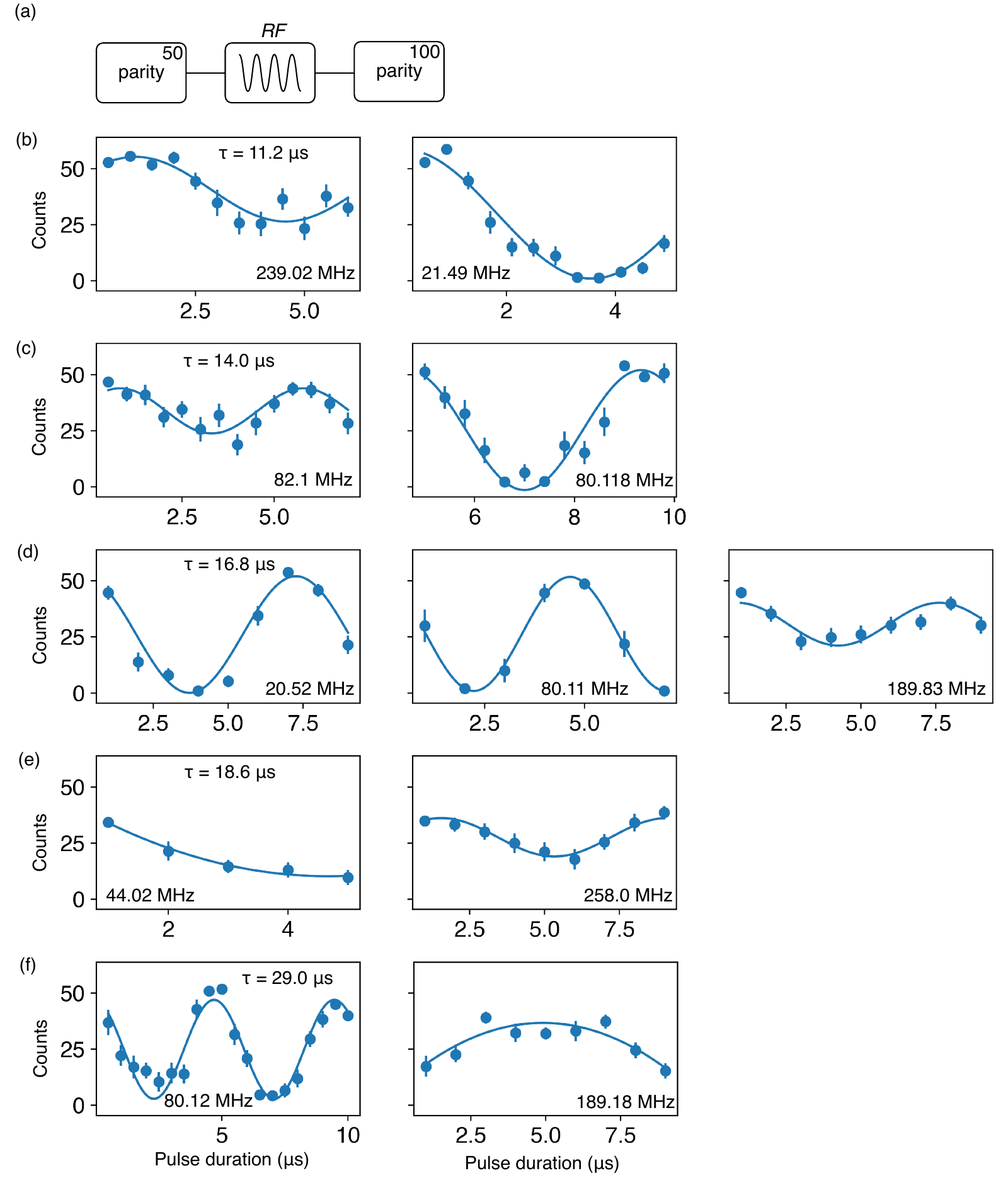}
\caption{\textbf{Rabi oscillations for different interpulse delays $\tau$.} \textbf{(a)} Experimental sequence. We use 50 parity readouts to initialize the spin pair in the antiparallel subspace by selecting on $\geq 15/50$ counts. The value of the interpulse delay $\tau$ in the parity readout determines which combination of Jahn-Teller and nitrogen-spin state we are initializing. Then, a radio-frequency (RF) pulse is applied. If resonant, the spin pair can flip to the parallel subspace and a Rabi oscillation is observed. \textbf{(b)} Results for $\tau = 11.2$ \textmugreek s. \textbf{(c)} Results for $\tau = 14.0$ \textmugreek s. \textbf{(d)} Results for $\tau = 16.8$ \textmugreek s. \textbf{(e)} Results for $\tau = 18.6$ \textmugreek s. \textbf{(f)} Results for $\tau = 29.0$ \textmugreek s. \textbf{(b-f)} The corresponding RF frequencies are given in the inset of each plot.}
\label{rf_rabi}
\end{figure}

To assign Jahn-Teller axes and nitrogen-spin states to signals observed at different values of the interpulse delay $\tau$, we measure Rabi oscillations and compare the frequencies at which signal was observed against Tables \ref{truth_table_A}, \ref{truth_table_B}, \ref{truth_table_C}, \ref{truth_table_D}. In Figure \ref{rf_rabi} we show the observed Rabi oscillations for each value of $\tau$. The contrast for $\tau = 18.6$ \textmugreek s is relatively poor, since there are other resonances close by (Fig. 2).

Due to mixing of the electron spin and nitrogen spin (Supplementary Sec. \ref{mixing}), the combinations of P1 eigenstates that can generate flip-flop dynamics can also include nitrogen spin flips. In particular, $\ket{0 \downarrow}$ and $\ket{- \uparrow}$ as well as $\ket{0 \uparrow}$ and $\ket{+ \downarrow}$ are mixed at our magnetic field and can therefore exhibit flip-flop dynamics. In Table \ref{possible_flip_flop_states} we show the Jahn-Teller axes and potential flip-flop states for each value of $\tau$ that we obtain from the combination of the Rabi oscillation experiments in Fig. \ref{rf_rabi} and Tables \ref{truth_table_A}, \ref{truth_table_B}, \ref{truth_table_C}, \ref{truth_table_D}. To evaluate which of the potential flip-flop states correspond to our observed dynamics, we enter all this information into the fit (Supplementary Sec. \ref{imaging}). Then, we obtain the fitted flip-flop states as shown in Table \ref{possible_flip_flop_states}.

\begin{table}[h]
\centering
\begin{tabular}{|c|c|c|c|}
\hline
$\tau$ (us) & Jahn-Teller axis & potential flip-flop states & fitted flip-flop states\\ \hline
11.2 & B & {$\ket{+ \uparrow}, \ket{+ \downarrow}$} or {$\ket{+ \downarrow}, \ket{0 \uparrow}$} & $\ket{+ \uparrow}, \ket{+ \downarrow}$\\ \hline
14.0 & A & {$\ket{0 \downarrow}, \ket{- \uparrow}$} or {$\ket{- \uparrow}, \ket{- \downarrow}$} & $\ket{- \uparrow}, \ket{- \downarrow}$\\ \hline
16.4 & A & {$\ket{0 \uparrow}, \ket{0 \downarrow}$} & $\ket{0 \uparrow}, \ket{0 \downarrow}$\\ \hline
18.6 & D & {$\ket{0 \uparrow}, \ket{+ \downarrow}$} or {$\ket{+ \uparrow}, \ket{+ \downarrow}$} & $\ket{+ \uparrow}, \ket{+ \downarrow}$\\ \hline
29.0 & B & {$\ket{0 \uparrow}, \ket{0 \downarrow}$} & $\ket{0 \uparrow}, \ket{0 \downarrow}$\\ \hline
\end{tabular}
\caption{\textbf{Potential and fitted flip-flop states for each measured $\tau$.} Each row indicates the Jahn-Teller axis, potential and fitted flip-flop states for the indicated value of $\tau$. The first index of the ket refers to the nitrogen spin, the second to the electron spin. When for example $\ket{+ \uparrow}, \ket{+ \downarrow}$ are the indicated basis states resulting in flip-flop dynamics, the corresponding P1-P1 eigenstates are $\frac{1}{\sqrt{2}}(\ket{+ \uparrow + \downarrow} \pm \ket{+ \downarrow + \uparrow})$.} 
\label{possible_flip_flop_states}
\end{table}

\clearpage

\section{Magnetic field fluctuations} \label{magnetic_field}

The external magnetic field (orientation) can change the effective P1-P1 electron-electron dipole interaction. The magnetic field fluctuations result from temperature fluctuations of the permanent magnets and from the presence of 6-9 T magnetic field systems in nearby laboratories. The effect of these fluctuations on the measured dipole-dipole interaction (Fig. \ref{X_distributions}) could ultimately limit the accuracy of the fitted P1-P1 position. To that end, we quantify the external magnetic field fluctuations by monitoring four single P1 frequencies using double electron-electron resonance (DEER). See Ref. \cite{Degen2021} for more details. 

\begin{figure}[h]
\includegraphics[width = \columnwidth]{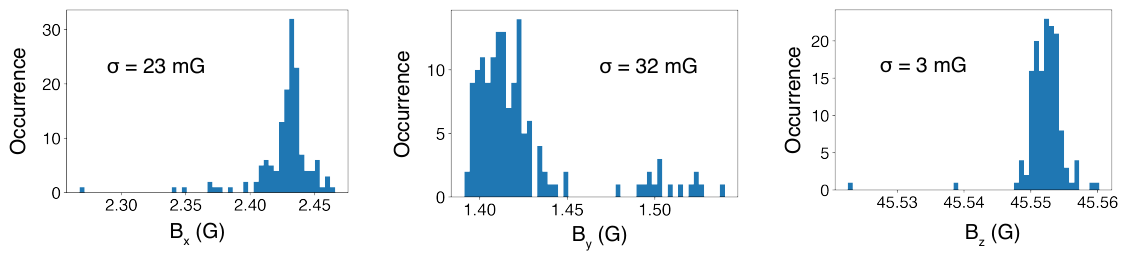}
\caption{\textbf{Magnetic field fluctuations during Ramsey experiments.} We plot the magnetic fields in the $x$, $y$ and $z$ direction during the Ramsey measurements in Fig. 3 of the main text. The magnetic fields are $B_x = 2.43(2)$ G, $B_y = 1.42(3)$ G and $B_z = 45.552(3)$ G. The standard deviation of the distribution $\sigma$ is given in the graphs. The components are obtained by measuring four single P1 frequencies using DEER. See ref. \cite{Degen2021} for more details.}
\label{B_distributions}
\end{figure}

\begin{figure}[h]
\includegraphics[width = 0.85\columnwidth]{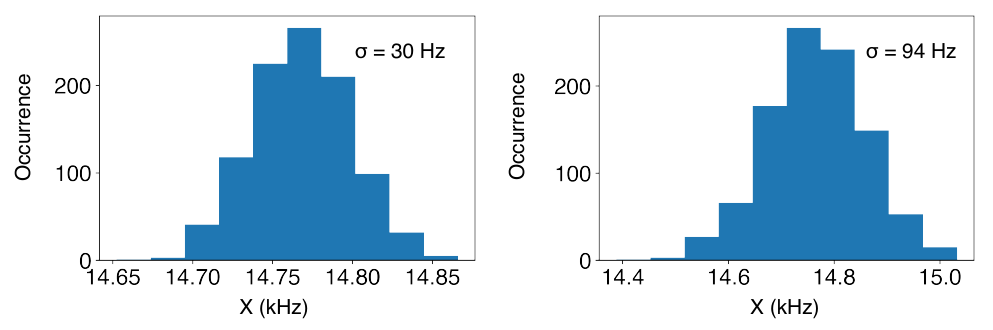}
\caption{\textbf{Effect of magnetic field fluctuations on P1-P1 electron-electron coupling.} We calculate the effective P1-P1 electron-electron dipolar coupling with the NV in $m_s = 0$ for various magnetic fields. Specifically, we take the obtained P1-P1 position and monitor the dipolar coupling when both P1 centers are in the Jahn-Teller state $A$ and the nitrogen-spin state $m_I = 0$. The magnetic field is $\mathbf{B} = [2.43(2), 1.42(3), 45.552(3)]$ G. On top of that, we add a random fluctuation on each value drawn from a Gaussian distribution with set standard deviations. (left) The fluctuations on $B_x$ and $B_y$ are 30 mG and the fluctuation on $B_z$ is 3 mG, consistent with $\sigma$ in Fig. \ref{B_distributions}. (right) The fluctuations on $B_x$ and $B_y$ are 100 mG and the fluctuation on $B_z$ is 20 mG, consistent with the approximate peak-to-peak values in Fig. \ref{B_distributions}. In this worst-case scenario we obtain an error on the dipolar coupling of $< 1 \%$.}
\label{X_distributions}
\end{figure}

The result is shown in Fig. \ref{B_distributions}. These are all the magnetic field measurements taken during the experimental period in which the Ramsey measurements in Fig. 3 of the main text were measured. Typical $B_x, B_y$ fluctuations are on the order of 30 mG and the $B_z$ fluctuation is 3 mG. The relative stability of $B_z$ is explained by the periodic recalibration of the magnetic field using the NV electron $m_s = 0$ to $m_s = -1$ frequency. However, during periods of the measurements, larger drifts are observed of about $100$ mG peak-to-peak in $B_x, B_y$ and $20$ mG in $B_z$. In both these regimes, we quantify the effect of such fluctuations on the measured dipolar coupling (Fig. \ref{X_distributions}). On average we find fluctuations with a standard deviation of $\sigma \sim 30$ Hz, which amounts to $0.2 \%$ relative to the dipolar coupling.

\clearpage

\section{Imaging the system} \label{imaging}
Combining the P1-P1 couplings obtained in this work with the NV-P1 couplings reported in previous work on this sample \cite{Degen2021}, we aim to resolve the spatial configuration of the NV-P1-P1 system. In our simulations, we construct a function that, given a magnetic field $\mathbf{B}$ and spatial configuration of the P1 centers, returns a set of couplings $\mathbf{C}$:
\begin{equation}
    f\left(\mathbf{B}, \mathbf{r}_{12},\mathbf{r}_{23}\right) = \mathbf{C}
    \label{eq:function}
\end{equation}
where 
\begin{table}[h]
    \centering
    \begin{tabular}{ll}
       $\mathbf{r}_{12}=[r_{12}, \theta_{12},  \phi_{12}]$  & \text{\hspace{1cm}vector from the NV center to the first P1 center} \\
        $\mathbf{r}_{23}=[r_{23}, \theta_{23},  \phi_{23}]$  & 
       \text{\hspace{1cm}vector from the first P1 center to the second P1 center}
    \end{tabular}
\end{table}\\
To resolve the physical position of the three defects, we use the least-squares fitting method \textit{scipy.optimize.leastq} which uses the Levenberg-Marquadt algorithm. Using the function in Eq. \ref{eq:function}, we provide a set of measured couplings $\mathbf{C}'$  and request a position that minimizes the residual sum of squares (RSS) between the calculated and measured coupling sets: $\mathbf{C}'$ and $\mathbf{C}$. The $\mathbf{B}$-field is a known and therefore fixed parameter. Finally, the errors on the obtained variables are the standard deviations. They are calculated from the covariance matrix of the variables returned from the fitting procedure.

\begin{figure}[b]
\centering
\includegraphics[width=0.5\linewidth]{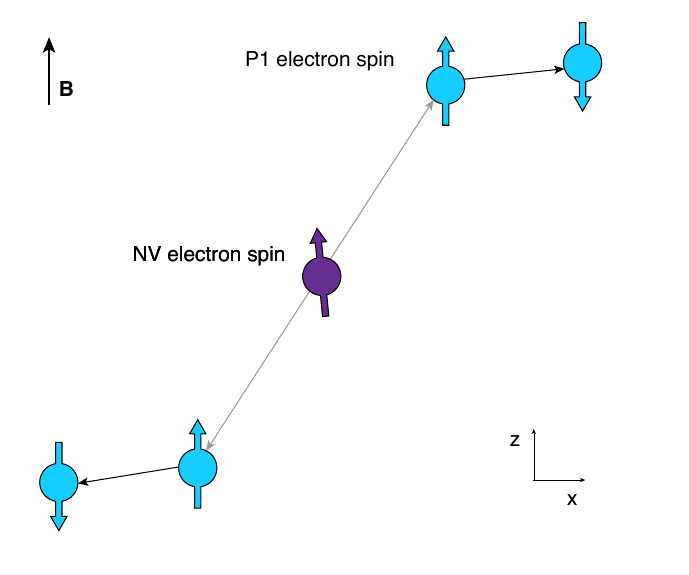}
    \caption{\textbf{Inversion symmetric solutions of the spatial configuration of the NV-P1-P1 system.} We find the relative position of one P1 center to the other. Due to the system's symmetry there are two possible solutions. For each, there is one unique solution for the relative position of both P1 centers to the NV. }
\label{fig:symmetric_solutions}
\end{figure}

We first find the relative position between the two P1 centers, which we label $S_1$ and $S_2$. We obtain two possible solutions. These are the two mirrored vectors corresponding to a permutation of the two P1 centers: $S_1 \xrightarrow{} S_2$ or $S_2 \xrightarrow{} S_1$. This symmetry is expected, due to the symmetry of the dipolar coupling. We then lock the P1-P1 position to one of the above vectors and find the position of the NV center with respect to the P1 pair. In total, this allows for two symmetrically inverted solutions. Figure \ref{fig:symmetric_solutions} shows a simplified schematic of the two possible solutions.

\subsection{Benchmarking}
To quantitatively analyse the performance of our imaging algorithm, we benchmark the fitting method in this section, for both the P1-P1 and NV-P1 pair, on numerically generated data for which the NV-P1-P1 positions are known. We generate $10$ random positions, calculate the exact couplings between the defects and introduce random errors on the couplings that reflect our measurement uncertainties. We sample these coupling errors from a normal distribution with a standard deviation of $0.2\%$, reflecting the standard deviation of the magnetic field fluctuations during our experiments in Supplementary Section \ref{magnetic_field}. We then apply our imaging algorithm and examine its performance by comparing the generated positions with the positions obtained through our fitting method. Due to multiple local minima of the function in Eq. \ref{eq:function}, the fitting results are sensitive to the initial guess. To tackle this, we fit each case with randomly generated initial guesses; $300$ and $400$ for the P1-P1 and NV-P1 pair systems respectively. Finally, we accept the outcome with the lowest RSS as the final position. 

\begin{figure}
\includegraphics[width=\textwidth]{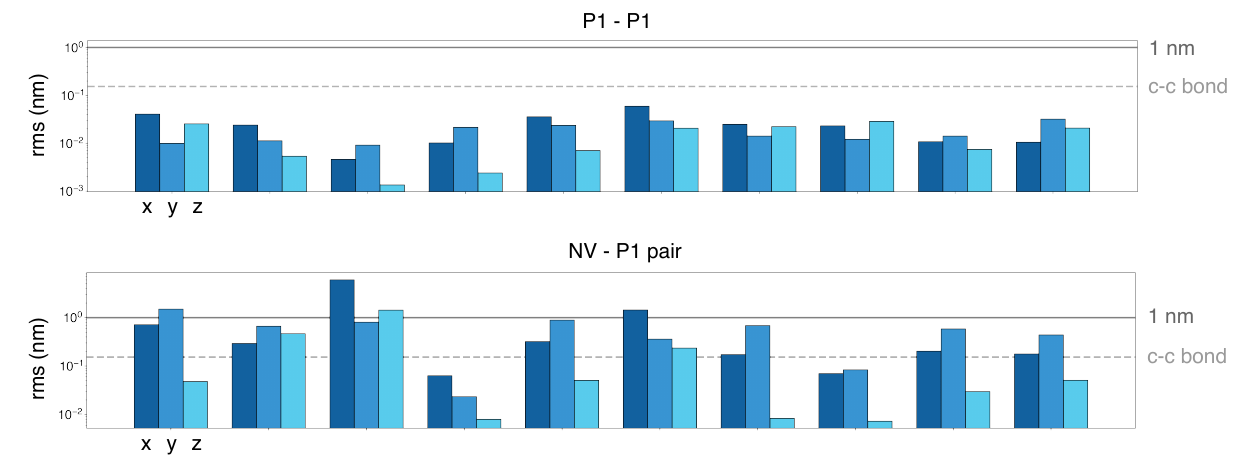}
    \caption{\textbf{Benchmarking of the fitting algorithm for the P1-P1 position and for the position of the NV center.} For the P1-P1 imaging benchmarking (top) we generate 10 random positions and calculate the exact couplings between them. We then create 200 different noisy coupling sets and execute our imaging process, with 300 initial guesses for each set. We accept the fit result with the lowest RSS value to the exact couplings and calculate the average deviation from the true position, in Cartesian coordinates. We repeat the same procedure to image the pair with respect to the NV center (bottom), where the P1 centers are explicitly set to their relative position, as obtained from the experimental couplings. We use 400 initial guesses. Note that although for the P1-P1 position, we achieve a resolution better than the diamond bond length, for the NV positions some errors exceed the nanometer mark.}
\label{fig:benchmarking}
\end{figure}

\subsection{Permutations}\label{sec:permutations}
\begin{figure}
    \centering
\includegraphics[width=0.8\linewidth]{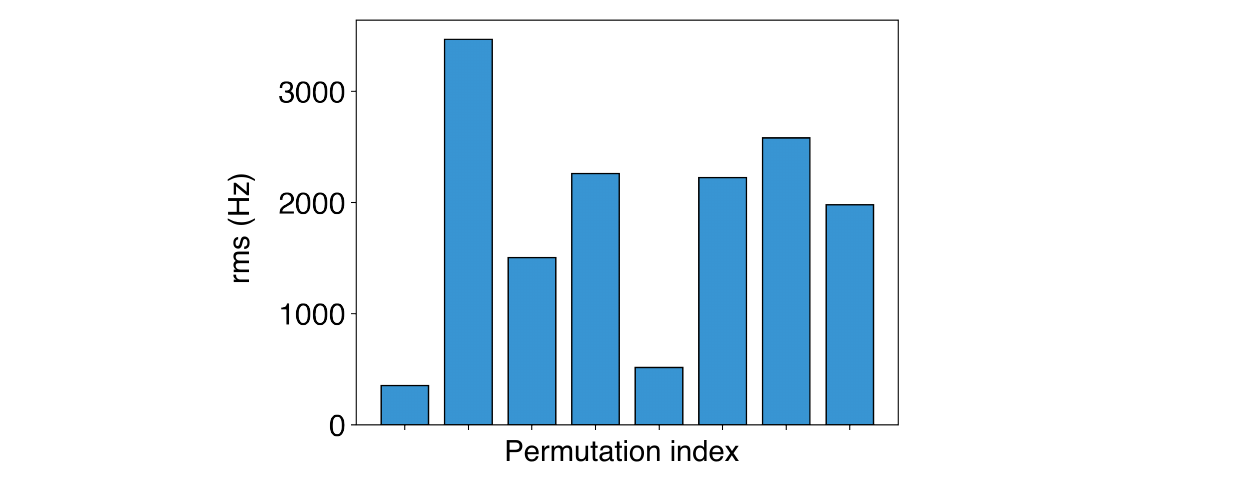}
    \caption{\textbf{Comparison of the RSS value for all possible nitrogen-spin state assignments.} We fit the experimental couplings for all the possible nitrogen-spin state combinations allowed by the RF measurements in Supplementary Section \ref{rabi}. The permutations are indexed according to Table \ref{tab:permutation_index}. The assignment with the lowest residual is the nitrogen-spin state assignment where the nitrogen spin is always the same for both P1 centers.}
    \label{fig:permutations}
\end{figure}

As discussed in Supplementary Section \ref{rf_simulations}, the RF measurements provide insight into the Jahn-Teller axis and nitrogen-spin state corresponding to a particular interpulse delay $\tau$. However, as discussed in Supplementary Section \ref{rabi}, we cannot uniquely identify the Jahn-Teller and nitrogen-spin state of the P1 pair directly from those measurements. Using the fitting algorithm described above, we consider all the possible states of the system, as indicated in Table \ref{possible_flip_flop_states}. Figure \ref{fig:permutations} shows the RSS values of the least-squares optimization method, for the $8$ possible permutations. The assignment with the lowest RSS value corresponds to the states where the nitrogen spin is fixed. For a more detailed discussion on the effect of electron-nitrogen spin mixing on the observed dynamical decoupling spectrum, see Supplementary Section \ref{mixing}. 

\begin{table}[h]
\centering
\resizebox{0.85\textwidth}{!}{%
\begin{tabular}{|c|c|c|c|c|c|}
\hline 
    Perm. Index & $\tau=11.2$ \textmugreek s ($B$) & $\tau=14.0$ \textmugreek s ($A$)& $\tau=16.4$ \textmugreek s ($A$) &  $\tau=18.6$ \textmugreek s ($D$) & $\tau=29.0$ \textmugreek s ($B$) \\\hline
   1 &$\ket{+\uparrow},\ket{+\downarrow}$ & $\ket{-\uparrow},\ket{-\downarrow}$ & $\ket{0\uparrow},\ket{0\downarrow}$& $\ket{+\uparrow},\ket{+\downarrow}$ & $\ket{0\uparrow},\ket{0\downarrow}$\\\hline
   2 &$\ket{+\uparrow},\ket{+\downarrow}$ & $\ket{-\uparrow},\ket{-\downarrow}$ & $\ket{0\uparrow},\ket{0\downarrow}$& $\ket{0\uparrow},\ket{+\downarrow}$ & $\ket{0\uparrow},\ket{0\downarrow}$\\\hline
   3 &$\ket{+\uparrow},\ket{+\downarrow}$ & $\ket{0\downarrow},\ket{-\uparrow}$ & $\ket{0\uparrow},\ket{0\downarrow}$& $\ket{+\uparrow},\ket{+\downarrow}$ & $\ket{0\uparrow},\ket{0\downarrow}$\\\hline
   4 &$\ket{+\uparrow},\ket{+\downarrow}$ & $\ket{0\downarrow},\ket{-\uparrow}$ & $\ket{0\uparrow},\ket{0\downarrow}$& $\ket{0\uparrow},\ket{+\downarrow}$ & $\ket{0\uparrow},\ket{0\downarrow}$\\\hline
   5 &$\ket{+\downarrow},\ket{0\uparrow}$ & $\ket{-\uparrow},\ket{-\downarrow}$ & $\ket{0\uparrow},\ket{0\downarrow}$& $\ket{+\uparrow},\ket{+\downarrow}$ & $\ket{0\uparrow},\ket{0\downarrow}$\\\hline
   6 &$\ket{+\downarrow},\ket{0\uparrow}$ & $\ket{-\uparrow},\ket{-\downarrow}$ & $\ket{0\uparrow},\ket{0\downarrow}$& $\ket{0\uparrow},\ket{+\downarrow}$ & $\ket{0\uparrow},\ket{0\downarrow}$\\\hline
   7 &$\ket{+\downarrow},\ket{0\uparrow}$ & $\ket{0\downarrow},\ket{-\uparrow}$ & $\ket{0\uparrow},\ket{0\downarrow}$& $\ket{+\uparrow},\ket{+\downarrow}$ & $\ket{0\uparrow},\ket{0\downarrow}$\\\hline
   8 &$\ket{+\downarrow},\ket{0\uparrow}$ & $\ket{0\downarrow},\ket{-\uparrow}$ & $\ket{0\uparrow},\ket{0\downarrow}$& $\ket{0\uparrow},\ket{+\downarrow}$ & $\ket{0\uparrow},\ket{0\downarrow}$\\\hline
\end{tabular}}
\caption{\textbf{The possible nitrogen-spin state assignments for all values of $\tau$.} We fit the measured couplings to all the possible combinations allowed by Tables \ref{truth_table_A}-\ref{truth_table_D}. The permutation index respects the order of Fig. \ref{fig:permutations}. The first entry of the denoted states refers to the nitrogen spin of the P1 center and the second to the electron spin. Each column indicates the possible flip-flop states for that particular interpulse delay $\tau$. The top of each column also indicates the corresponding Jahn-Teller axis in brackets.}
\label{tab:permutation_index}
\end{table}

\clearpage

\section{Simulation of dynamical decoupling spectrum} \label{simulation}

The NV-P1-P1 positions obtained in the main text allow us to simulate the dynamical decoupling spectrum measured in Fig. 2d. The result is shown in Fig. 2e. We observe good qualitative agreement between the simulated dynamical decoupling spectrum and the measured spectrum.

To simulate the spectrum, we have to consider all possible Jahn-Teller axes and P1-P1 eigenstates. Therefore, we simulate the dynamical decoupling signal of the NV center electron spin for each Jahn-Teller axis and for each corresponding eigenstate of the P1-P1 Hamiltonian. We then convert the simulated fidelity to an observable number of counts. When the NV center electron spin is in $m_s = 0$, the probability of measuring a photon count is $70 \%$. And when the electron spin is in $m_s = -1$, the probability of not measuring a photon count is $99 \%$. Given the 200 repetitive dynamical decoupling repetitions, we can thus convert the simulated NV electron spin fidelity to the experimental number of counts. Finally, we add Poissonian noise on the photon counts to simulate shot noise.

Given the expected signal per Jahn-Teller axis and corresponding eigenstate, we now assume each Jahn-Teller axis and eigenstate has equal probability of occurrence. Then, we sum all the individual signals and normalise the result. This procedure results in the simulated dynamical decoupling signal in Fig. 2e.

In the simulated spectrum, we observe signal originating from flip-flop states that only involve the electron spin, but we also observe signal from the more complex flip-flop states that involve both the nitrogen and the electron spin. The latter signals mainly occur at $\tau = 20-25$ \textmugreek s (Supplementary Sec. \ref{mixing}). In the experimental data (Fig. 2d) we also see clear signal at these values of $\tau$, indicating that we observe flip-flop dynamics involving the two nitrogen spins in experiment.

\section{Ramsey data with the NV electron spin in $m_s = -1$} \label{ramsey}

In the main text, we perform Ramsey experiments using five values of the interpulse delay $\tau$ (Fig. 3). For these measurements, the NV electron spin is in $m_s = 0$ during free evolution. This is done to isolate the P1-P1 dipolar coupling from the NV-P1 dipolar couplings. In that way, we can fit to the P1-P1 position first before bringing in the NV electron spin.

\begin{figure}
\centering
\includegraphics[width = 0.55\columnwidth]{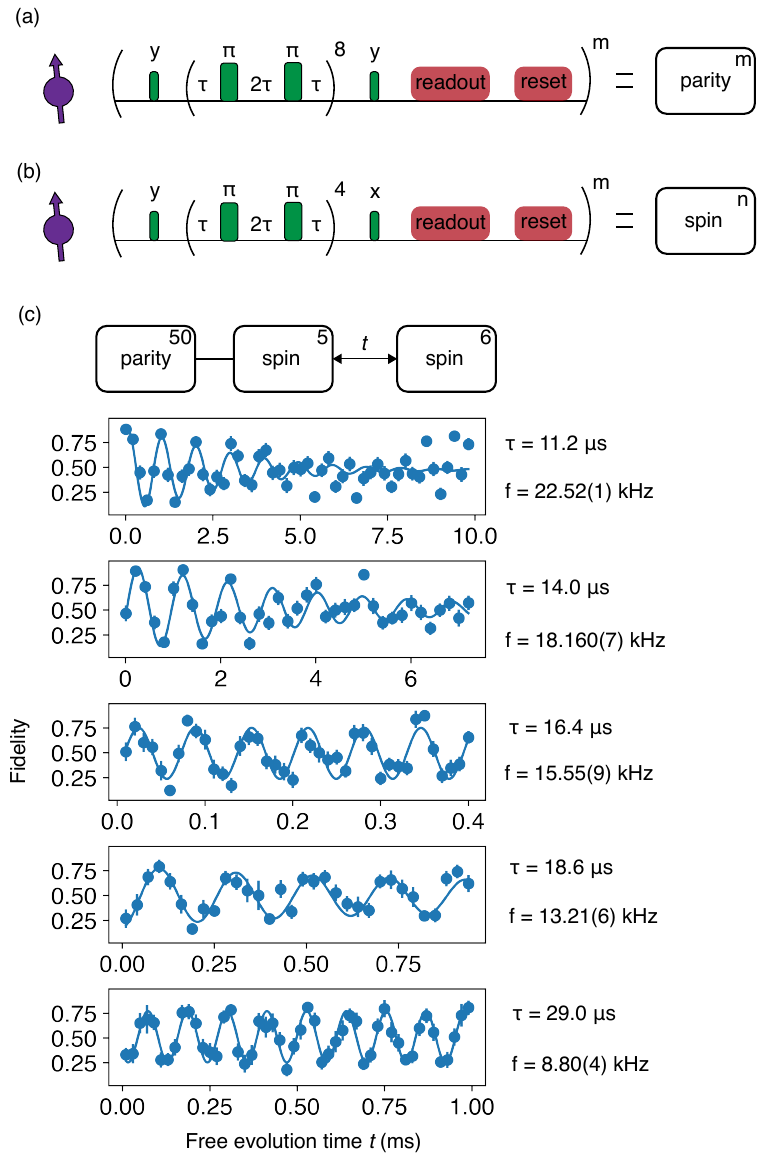}
\caption{\textbf{Ramsey measurements of the P1 electron spin pair when the NV electron spin is in $m_s = -1$.} \textbf{(a,b)} Pulse sequences to measure the parity ($\{\ket{\uparrow \uparrow},\ket{\downarrow \downarrow}\}$ vs. $\{\ket{\uparrow \downarrow},\ket{\downarrow \uparrow}\}$) and spin ($\{\ket{\uparrow \downarrow}\}$ vs. $\{\ket{\downarrow \uparrow}\}$) of the electron-spin pair. \textbf{(c)} Ramsey measurements for five different interpulse delays $\tau$. Each $\tau$ corresponds to an identifiable dip in the dynamical decoupling spectrum of Fig. 2d. The obtained frequency $f$ as well as the interpulse delay $\tau$ used are indicated on the right.}
\label{ramsey_ms1}
\end{figure}

In Fig. \ref{ramsey_ms1} we show the Ramsey experiments for each of the five values of $\tau$ with the NV electron spin in $m_s = -1$. Next to the major contribution of the P1-P1 dipolar coupling, we also get a contribution of the NV-P1 dipolar coupling, which adds a detuning to the P1-P1 coupling and thereby alters the measured flip-flop frequency. In Table \ref{ramsey_ms1_comparison} we compare the measured evolution frequencies of the Ramsey experiment to the expected values from the NV-P1-P1 position found in the main text. Overall we find good agreement, but we do observe deviations of up to a few hundred Hz. In particular, the deviations for $\tau = 16.4$ \textmugreek s and $\tau = 18.6$ \textmugreek s with the NV electron spin in $m_s = -1$ are larger than expected. Currently we cannot explain these deviations.

An additional observation from Fig. \ref{ramsey_ms1} is that the inhomogeneous dephasing times with the NV electron spin in $m_s = -1$ are about an order of magnitude smaller than those with the NV electron spin in $m_s = 0$. This is expected, since the additional field gradient due to the presence of the NV electron spin detunes the P1 electron-spin pair away from the anticrossing making it more susceptible to magnetic field noise (i.e. a less effective clock transition).

\begin{table}[h]
\small
\centering
\begin{tabular}{|c|c|c|c|c|}
\hline
$\tau$ (us) & measured $f_{m_s = -1}$ (kHz) & calculated $f_{m_s = -1}$ (kHz) & measured $f_{m_s = 0}$ (kHz) & calculated $f_{m_s = 0}$ (kHz)\\ \hline
11.2 & 22.52(1) & 22.378 & 22.152(2) & 22.106 \\ \hline
14.0 & 18.160(7) & 18.323 & 17.943(1) & 18.114 \\ \hline
16.4 & 15.55(9) & 15.027 & 14.87(8) & 14.837 \\ \hline
18.6 & 13.21(6) & 13.856 & 13.7(1) & 13.414 \\ \hline
29.0 & 8.80(4) & 8.892 & 8.2(3) & 8.591 \\ \hline
\end{tabular}
\caption{\textbf{Measured and calculated frequencies of the P1 electron-spin pair.} For each $\tau$, we indicate the measured frequency when the NV electron spin is in $m_s = -1$ as well as the calculated frequency based on the obtained position in the main text. For completeness, we also give the frequencies when the NV electron spin is in $m_s = 0$.} 
\label{ramsey_ms1_comparison}
\end{table}

\clearpage

\section{Dephasing mechanisms} \label{dephasing}

In this section, we discuss the mechanisms involved in the P1-P1 electron-spin pair $T_2^*$. The spin-pair dephasing originates from magnetic field fluctuations, either from the permanent magnets used to apply our magnetic field or from the local spin baths surrounding the spin pair. The electron-spin pairs in our diamond are surrounded by two different spin baths:
\begin{itemize}
    \item $^{13}$C spin bath with a concentration of $\sim 0.01 \%$ \cite{Bradley2022}
    \item P1 electron-spin bath with an estimated concentration of $\sim 75$ ppb \cite{Degen2021}
\end{itemize}

The combined noise of the external magnetic field and the two local spin baths can have a different effect on the two P1 centers forming the pair. We will denote the magnetic field noise on the first (second) P1 center electron spin as $\delta B_1$ ($\delta B_2$). These noise terms affect the evolution frequency of the spin pair, which results in dephasing. To solve this generally, we have to consider the full effect of $\delta B_{1,2}$ on the P1-P1 Hamiltonian, including the time dependence of $\delta B_{1,2}$. Here, we will assume that $\delta B_{1,2}$ is quasi-static, which is not necessarily true for the bath of P1 centers. Furthermore, we will consider the effects of global, correlated noise and local, uncorrelated noise separately.

\subsection{Correlated and uncorrelated noise}

The quantisation axis of the P1 center is generally not along the direction of the external magnetic field due to the strong electron-nuclear hyperfine interaction ($\gamma_e B \sim A_\parallel, A_\perp$). Correlated noise, such as fluctuations of the global magnetic field strength, therefore has a non-negligible effect on the effective coupling $X$ between the two electron spins. In contrast, $^{13}$C nuclear spin pairs form a more ideal decoherence-free subspace, since fluctuations of the global magnetic field strength do not influence the nuclear-nuclear effective coupling at all \cite{Bartling2022}. On the other hand, uncorrelated, local noise from the spin bath affects the detuning of the two P1 center electron spins. In the pseudo-spin picture, we can write this uncorrelated noise $\Delta Z$ as \cite{Bartling2022}:

\begin{equation} \label{pseudo_spin_noise}
    H = X \hat{S}_x + [m_s Z + \Delta Z] \hat{S}_z.
\end{equation}

\begin{figure}[h]
\includegraphics[width = 0.9\columnwidth]{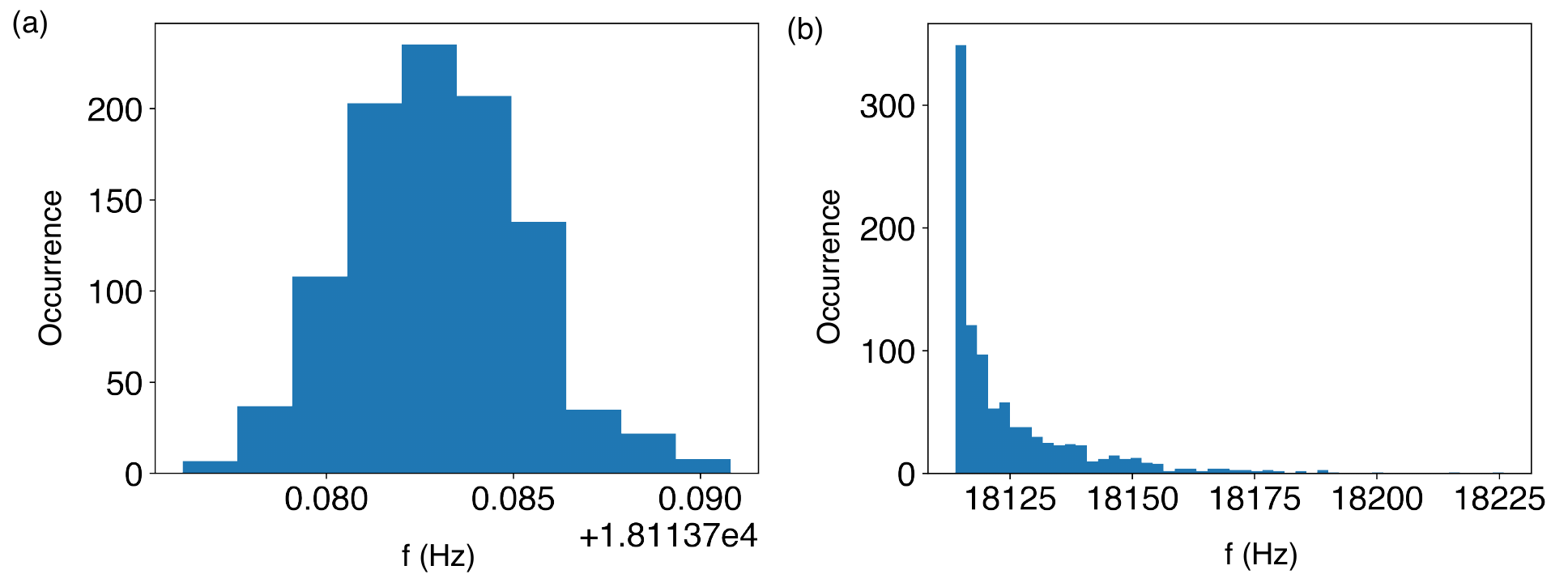}
\caption{\textbf{Simulation of the effect of correlated and uncorrelated noise of the nuclear-spin bath on the two P1 centers at $\tau = 14.0$ \textmugreek s.} We simulate the P1-P1 system discussed in this paper for Jahn-Teller axis $A$ and nitrogen-spin state $m_I = -1$, which corresponds to the resonance at $\tau = 14.0$ \textmugreek s. Then, we calculate the effective electron-electron coupling for two different situations. \textbf{(a)} We add noise with a standard deviation of $\sigma = 0.3$ mG to the $z$-direction of the external magnetic field. The value of 0.3 mG corresponds to the estimated noise due to the $^{13}$C bath. In this simulation, we assume the noise is common to both P1 centers and therefore correlated. Therefore, its main effect is to modulate the effective electron-electron coupling. \textbf{(b)} We vary the magnetic field in the $z$-direction of only one of the P1 centers. The standard deviation is also $\sigma = 0.3$ mG. This simulation emulates the noise from nearby nuclear spins that have a different effect on each of the P1 centers forming the pair. The noise is therefore uncorrelated.}
\label{anticrossing}
\end{figure}

To examine the difference between the effect of correlated (global) noise and uncorrelated (local) noise on a P1-P1 electron-spin pair, we simulate the P1-P1 system obtained in the main text for the resonance at $\tau = 14.0$ \textmugreek s. We consider correlated noise to be the exact same noise on both P1 centers, also in magnitude, and uncorrelated noise to be completely independent noise on each P1 center. To simulate an example of correlated noise, we vary the $z$-component of the external magnetic field with a standard deviation of $\sigma = 0.3$ mG (Fig. \ref{anticrossing}a). To simulate uncorrelated noise, we vary the $z$-component of the external magnetic field for only one of the two P1 centers (Fig. \ref{anticrossing}b). From the results in Fig. \ref{anticrossing}, we conclude that the effect of correlated noise on the spin-pair frequency is negligible compared to the effect of uncorrelated noise. 

Additionally, the two types of noise result in different distributions, which can be understood from the different origins of noise. Correlated noise does not affect the detuning $Z$ between the two spins, but it does affect the coupling $X$. The noise therefore adds linearly to $X$, leading to a relatively symmetric distribution. On the other hand, uncorrelated noise adds quadratically to the frequency, as is shown in Equation \ref{pseudo_spin_noise}. This results in a one-sided distribution, as the effect of negative and positive noise $\Delta Z$ is the same. Figure \ref{anticrossing}b also highlights that the spin pair is only second-order sensitive to uncorrelated noise for $m_s = 0$ and therefore forms a clock transition.

For similar magnitudes of correlated and uncorrelated noise, the uncorrelated noise dominates in limiting the $T_2^*$ of a spin pair. In other words, it is the difference between $\delta B_1$ and $\delta B_2$ that determines $\Delta Z$, which is the main contributor to the inhomogeneous dephasing of the spin pair.

\begin{table}
\footnotesize
\begin{tabular}{|p{3cm}|p{4cm}|p{3cm}|p{2.5cm}|p{3cm}|}
 \hline
 \text{Noise source} & \text{Magnitude} & \text{Type} & \text{Pseudo-spin effect} & \text{Expected $T_2^*$} \\  \hline
 Nuclear-spin bath & 0.89(4) kHz & Mostly uncorrelated & $X,Z$ & $\sim 10-40$ ms \\  \hline
 P1 spin bath & 2.22(5) kHz & Uncorrelated \newline\& correlated & $X,Z$ & Uncorrelated: $\sim 5$ ms \newline Correlated: $\gg 1$ s  \\ \hline
 External B-field & $[\sigma_x, \sigma_y, \sigma_z] = [23, 32, 3]$ mG \newline $[\sigma_x, \sigma_y, \sigma_z] = [3, 3, 3]$ mG & Correlated & $X$ & $\sim 7.5$ ms \newline $\sim 83$ ms  \\ \hline
\end{tabular}
\caption{\textbf{Summary of noise sources and their effects.} For each noise source, we summarise their magnitude, type (correlated or uncorrelated), expected $T_2^*$ and whether that noise source mainly affects the coupling between the spins $X$ or whether it affects both the coupling $X$ and the detuning between the two spins $Z$.}
\label{noise_table}
\end{table}

\subsection{Nuclear-spin bath}
To estimate the typical noise strength from both the nuclear-spin bath and the electron-spin bath, we consider the NV electron spin $T_2^*$, which is measured to be $94(2)$ \textmugreek s \cite{Bradley2022}. Assuming quasi-static, Gaussian noise this gives a standard deviation of the frequency of $\sigma_f = \frac{1}{\sqrt{2} \pi T_2^*} = 2.39(5)$ kHz. 

\begin{figure}
\includegraphics[width = 0.9\columnwidth]{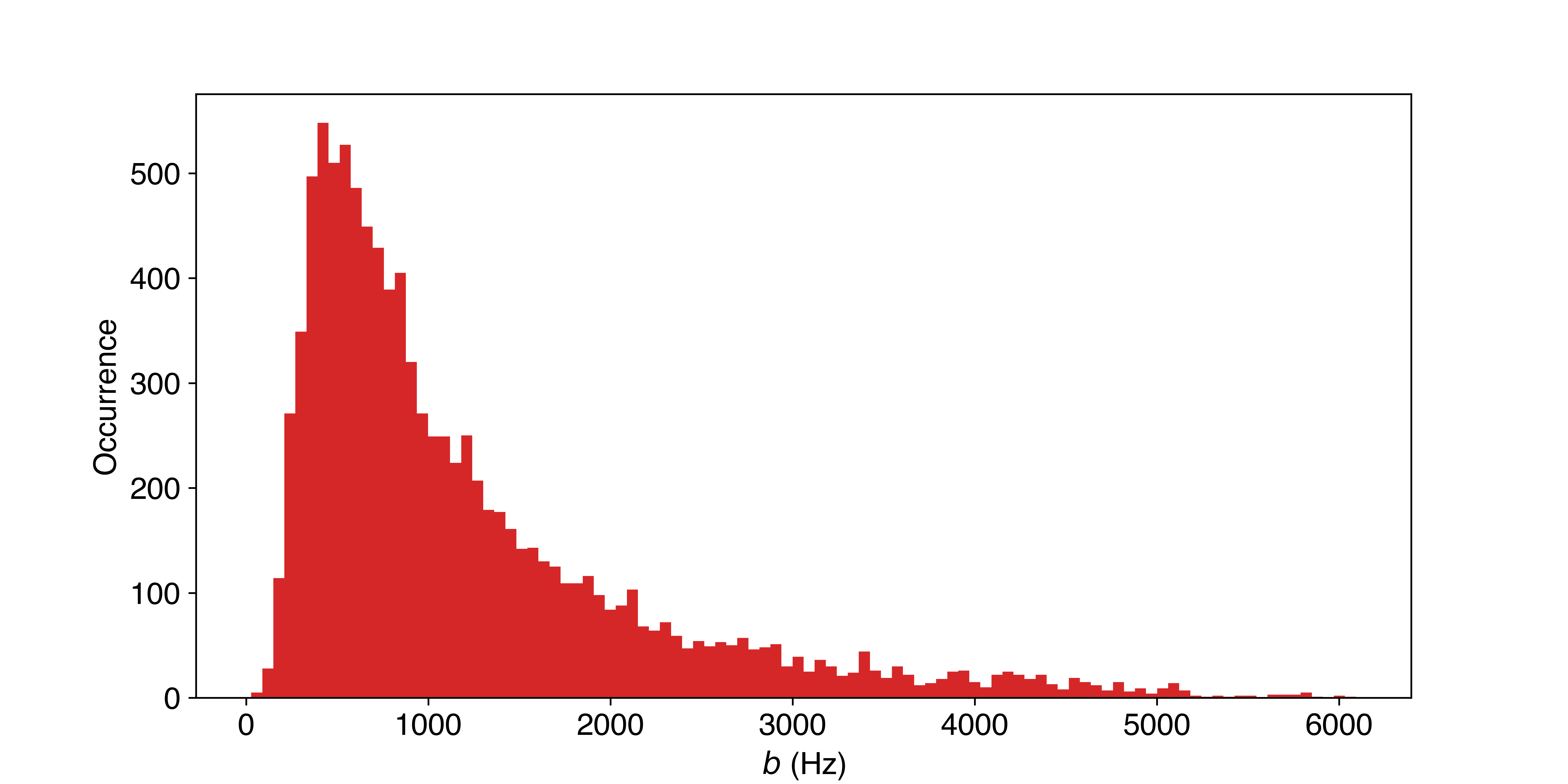}
\caption{\textbf{Simulated electron-spin noise for varying $^{13}$C spin configurations.} We simulate $10^4$ configurations of $^{13}$C spins surrounding an electron spin in a diamond lattice for a concentration of 0.01 $\%$. The standard deviation of the noise generated by one such $^{13}$C spin bath is $b$. The average of the distribution is 1.2 kHz. The right tail is due to more strongly coupled spins (we excluded spins with a coupling larger than 10 kHz).}
\label{carbon_noise}
\end{figure}

To estimate what part of this noise is typically due to the $^{13}$C bath, we follow Ref. \cite{Bradley2022} where the $T_2^*$ of a $^{13}$C spin due to the $^{13}$C bath in the same device was measured to be $0.66(3)$ s. This gives $\sigma_f = 0.34(2)$ Hz. To convert this to the noise on the electron spin, we multiply by $\frac{\gamma_e}{\gamma_c}$ where $\gamma_e$ ($\gamma_c$) is the electron ($^{13}$C) gyromagnetic ratio. We obtain $\sigma_f = 0.89(4)$ kHz. As an alternative approach, we simulate $10^4$ configurations of $^{13}$C spins surrounding an electron spin with a concentration of $0.01 \%$. The result is shown in Fig. \ref{carbon_noise}. The average noise an electron spin experiences due to a $^{13}$C bath is 1.2 kHz, similar to the value of $0.89(4)$ kHz obtained from the measurement in Ref. \cite{Bradley2022}. Note that the exact values for each spin depend strongly on the local environment, and therefore these numbers should only be interpreted as estimates for typical values of the standard deviation of the noise and its distribution.   

Importantly, the $^{13}$C spins are relatively close to the P1 centers. Therefore, the noise due to the $^{13}$C spins on both P1 centers of the pair is likely uncorrelated (local), although we cannot quantify how uncorrelated it is exactly. Since the effect of uncorrelated noise relative to correlated noise is large, we can follow the analyses of Dobrovitski et al. \cite{Dobrovitski2009} and Bartling et al. \cite{Bartling2022} to analyse Equation \ref{pseudo_spin_noise}. We plot the analytical solution in Fig. \ref{p1_coherence_analytical}a for a quasi-static bath with a noise magnitude of 0.3 mG and a coupling $X = 18.114$ kHz. We roughly reproduce the timescales of the observed decay, suggesting that the uncorrelated noise from the $^{13}$C spin bath plays an important role in limiting the P1 pair inhomogeneous dephasing time. Note that we have assumed that all estimated $^{13}$C bath noise is uncorrelated, while it is conceivable it is partially correlated. That would increase the inhomogeneous dephasing time in Fig. \ref{p1_coherence_analytical}a. 

\subsection{Electron-spin bath}
\begin{figure}
\includegraphics[width = 0.88\columnwidth]{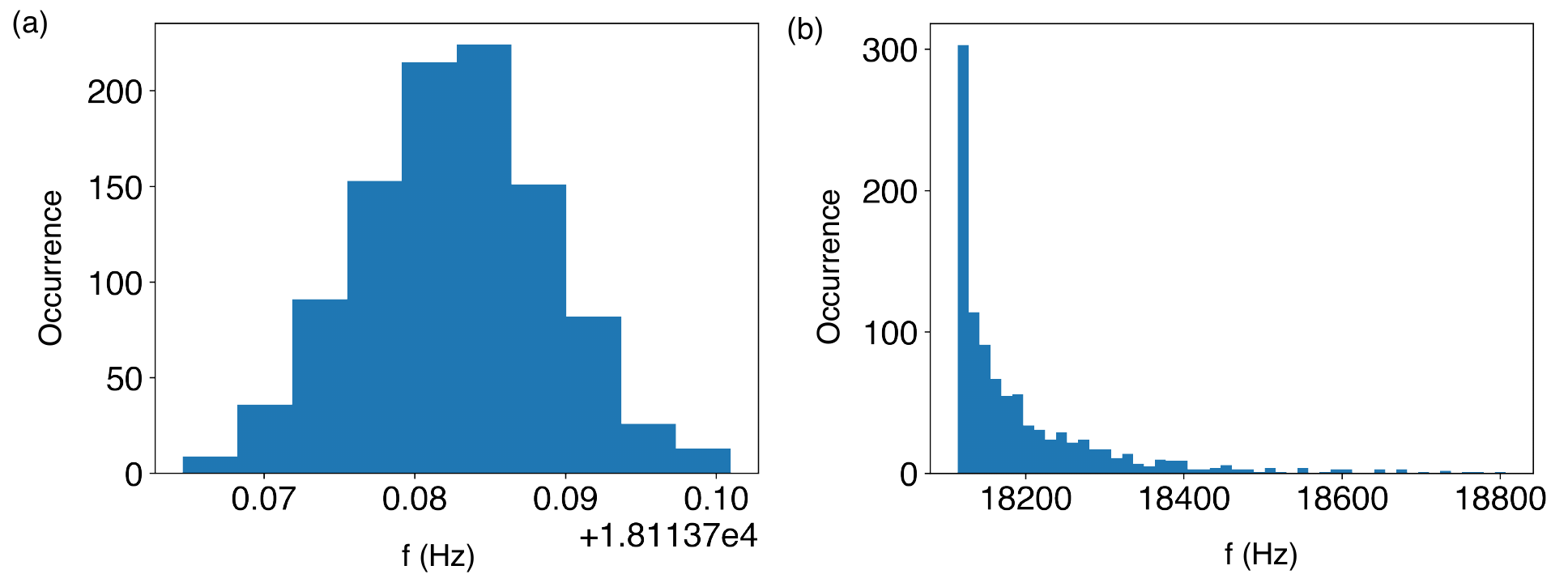}
\caption{\textbf{Simulation of the effect of correlated and uncorrelated noise of the P1 bath on the two P1 centers at $\tau = 14.0$ \textmugreek s.} We simulate the P1-P1 system discussed in this paper for Jahn-Teller axis $A$ and nitrogen-spin state $m_I = -1$, which corresponds to the resonance at $\tau = 14.0$ \textmugreek s. Then, we calculate the effective electron-electron coupling for two different situations. \textbf{(a)} We add noise with a standard deviation of $\sigma = 0.8$ mG to the $z$-direction of the external magnetic field. The value of 0.8 mG corresponds to the estimated noise due to the P1 bath. This noise is common to both P1 centers and therefore correlated. Therefore, its main effect is to modulate the effective electron-electron coupling. \textbf{(b)} We vary the magnetic field in the $z$-direction of only one of the P1 centers. The standard deviation is also $\sigma = 0.8$ mG. This simulation emulates the noise from nearby electron spins that have a different effect on each of the P1 centers forming the pair. The noise is therefore uncorrelated.}
\label{anticrossing_p1}
\end{figure}

\begin{figure}
\includegraphics[width = 0.92\columnwidth]{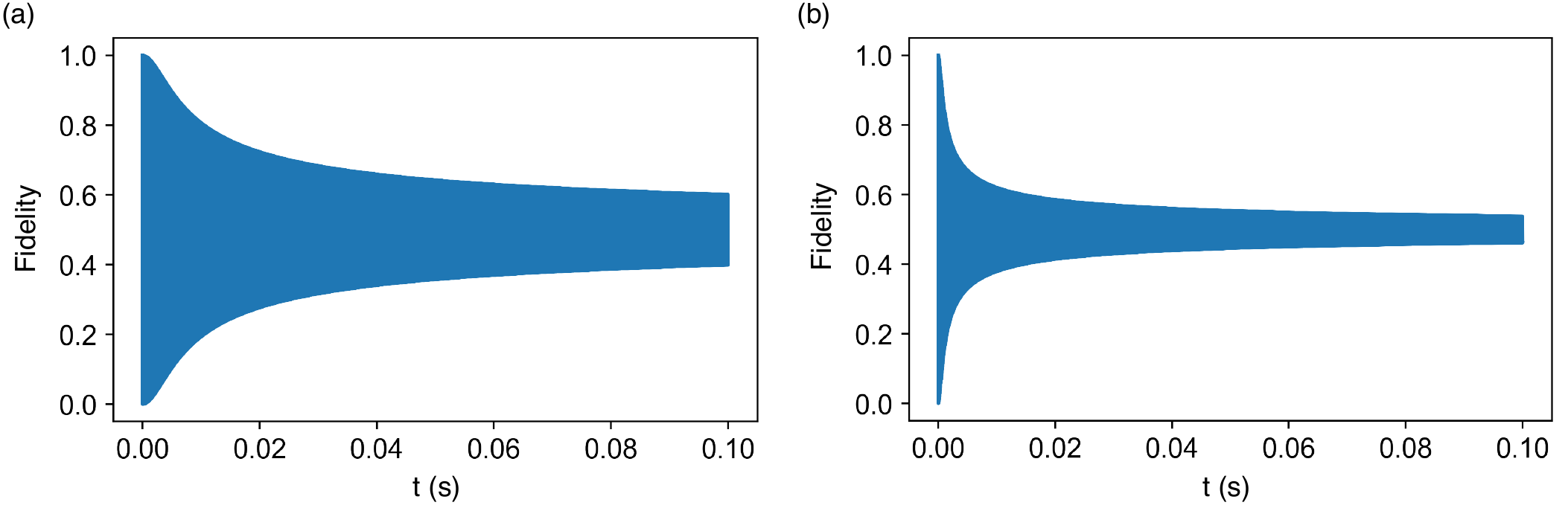}
\caption{\textbf{The expected P1 pair inhomogeneous dephasing under the assumption of uncorrelated noise.} We plot the analytical solutions to Equation \ref{pseudo_spin_noise} obtained from Refs. \cite{Dobrovitski2009,Bartling2022}. \textbf{(a)} The spin-pair coupling is $X = 18.114$ kHz and the standard deviation of the noise is 0.3 mG, consistent with the expected noise from the nuclear-spin bath. We find a timescale comparable to the experimentally observed $T_2^* = 44(9)$ ms. \textbf{(b)} With a noise of 0.8 mG, consistent with the noise from the electron-spin bath under the assumption that all P1 bath noise is uncorrelated and quasi-static.}
\label{p1_coherence_analytical}
\end{figure}

We estimate the noise from the electron spin bath on a single electron spin considering the noise on the NV center ($2.39(5)$ kHz) and the nuclear-spin bath noise ($0.89(4)$ kHz): $b_{\text{electron}} = \sqrt{b_{\text{total}}^2 - b_{\text{nuclear}}^2} = 2.22(5)$ kHz. This noise figure consists of a correlated and an uncorrelated part on the pair of P1 centers. We do not know exactly what part of the $2.22(5)$ kHz noise is correlated and what part is uncorrelated. In Fig. \ref{anticrossing_p1} we show the effect of the P1 bath noise either being fully correlated or fully uncorrelated. When we assume the noise to be fully correlated, we obtain a modulation of the coupling $X$ with a standard deviation of $6$ mHz, resulting in $T_2^*$ values exceeding a second. 

When the noise is uncorrelated, we can perform the same analysis as for the nuclear spins. We follow the analyses of Dobrovitski et al. \cite{Dobrovitski2009} and Bartling et al. \cite{Bartling2022} to analyse Equation \ref{pseudo_spin_noise}. This results in the dephasing as shown in Fig. \ref{p1_coherence_analytical}b. We observe a decay time of a few milliseconds. Note that in this analysis we assume that the P1 bath is quasi-static, which is likely not a valid assumption for longer times.

\subsection{External magnetic field}
The long-term magnetic field fluctuations are $\sigma_x = 23$ mG, $\sigma_y= 32$ mG and $\sigma_z = 3$ mG. In Figure \ref{X_distributions} we simulate the effect of external magnetic field fluctuations on the electron-electron coupling. Due to the relative magnitude of the $\sigma_x$ and $\sigma_y$ fluctuations, a significant standard deviation of $\sigma = 30$ Hz on the electron-electron coupling is obtained. This translates to $T_2^* \approx 7.5$ ms, smaller than the observed value of $T_2^* = 44(9)$ ms. 

For the Bloch vector measurements in Fig. 3d, we measure $\langle Y \rangle$ and $\langle Z \rangle$ for each point after which we obtain the Bloch vector length as $\sqrt{\langle Y \rangle^2 + \langle Z \rangle^2}$. Slow magnetic field fluctuations over the course of the measurement do not affect the Bloch vector measurement as much as a Ramsey measurement. Only the relative phase between $\langle Y \rangle$ and $\langle Z \rangle$ within a single data point is prone to fluctuations, but the Bloch vector measurement is not sensitive to different external magnetic fields between data points, contrary to a Ramsey measurement.

\begin{figure}
\centering
\includegraphics[width = 0.4\columnwidth]{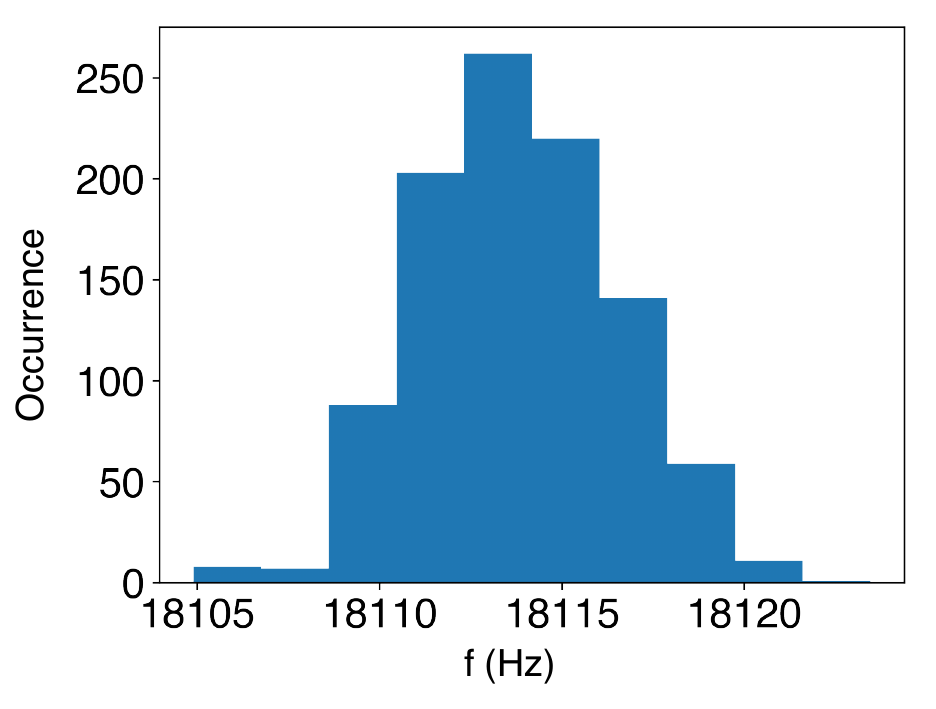}
\caption{\textbf{Simulation of the effect of correlated and uncorrelated noise of the external magnetic field on the two P1 centers at $\tau = 14.0$ \textmugreek s.} We simulate the P1-P1 system discussed in this paper for Jahn-Teller axis $A$ and nitrogen spin state $m_I = -1$, which corresponds to the resonance at $\tau = 14.0$ \textmugreek s. Then, we calculate the effective electron-electron coupling for a standard deviation of $\sigma = 3.0$ mG on the $x$-, $y$- and $z$-direction of the external magnetic field. This noise is common to both P1 centers and therefore correlated. Therefore, its main effect is to modulate the effective electron-electron coupling.}
\label{anticrossing_B}
\end{figure}

In our experiment, the external magnetic field fluctuations are typically much slower than the duration of a single measurement point. If we then assume more conservative fluctuations during the Bloch vector measurement of $\sigma_x = \sigma_y = \sigma_z = 3$ mG, we observe an effect on the coupling $X$ as shown in Fig. \ref{anticrossing_B}. The standard deviation on the coupling $X$ is 2.7 Hz, which translates to $T_2^* = 83$ ms. \\

We conclude that all three noise sources can have a significant effect on the observed inhomogeneous dephasing time of $T_2^* = 44(9)$ ms. In Table \ref{noise_table} we summarise the effects of the three noise sources.

For the nuclear- and electron-spin baths, the correlated noise is negligible compared to the uncorrelated noise. The nuclear-spin bath noise is likely more uncorrelated due to their closeness to the P1 centers. However, it is unclear exactly what part of the noise is correlated and what part is uncorrelated. To obtain a more thorough description of the spin-bath noise, the complex dynamics of the P1 bath would have to be taken into account using for example correlated cluster expansion (CCE). 

The magnetic field fluctuations are larger in magnitude, but only introduce correlated noise. The Bloch vector length measurement is crucial to mitigate the longer-time fluctuations of the external magnetic field.

\clearpage

\section{Optimization of parity initialization} \label{optimization}
We perform measurement-based initialization: measurements are performed to confirm that the system is currently in the desired state (JT, $^{14}$N spin and spin state for both P1 centers). Due to the many possible states the P1 pair can take, the probability to start in the subspace is approximately 1/288, making initialization time-consuming. In this section, we describe how we optimize the initialization procedure for speed and fidelity, by using repeated measurements and intermediate dynamic decision making combined with the capability to scramble the states. The factor $\sim 10$x speed-up realized is essential for enabling the experiments in the main text.  

To initialize the P1 electron spin pair, we execute $m$ parity measurements, followed by $n$ spin measurements. The parity measurement initializes the two electron spins of the P1 centers into the antiparallel subspace and the desired $^{14}$N and JT configuration. The spin measurement initializes in one of the two antiparallel spin states. If the P1 centers are in a particular configuration, such that their coupling is resonant with the dynamical decoupling interpulse delay, the NV electron is projected into the bright state and we detect photons (Fig. 2). To find a robust initalization scheme, we implement repetitive parity measurements, at some specific interpulse delay $\tau$, and obtain a time-trace for the pair, as shown in Fig. 2b. By collapsing the time-trace in bins of $200$ parity measurements we make two observations (Fig. 2c). We note a well-separated peak between 120 and 150 counts. This allows us to introduce a threshold check for initialization, which we implement with $50$ parity measurements. If we observe more than 15 photons in $50$ parity measurements, the pair is initialized in the Jahn-Teller and nitrogen spin configuration resonant to the used interpulse delay $\tau$, and in the anti-parallel electron spin state.

To speed up the initialization time, we introduce intermediate photon count thresholds. We check after a number of parity measurements whether we already have observed photon counts, and decide whether we want to abort and restart the initialization. For example, consider a total of $50$ parity measurements with a threshold of $15$ photon counts. If we perform an intermediate check at the $30$th iteration and we have not detected any photons, it is highly unlikely to detect $15$ photons at the $50$th parity measurement. In this scenario, we abort the sequence early. We randomize the Jahn-Teller axes, and nitrogen-spin states of the P1 centers by applying a green laser pulse and restart the initialization procedure \cite{Degen2021}. 

We optimize for minimum initialization time by analyzing data sets such as in Fig. 2b with a Monte Carlo sampling method. By starting at a random point along the time trace, we emulate parity measurements of P1 pairs that are in a random Jahn-Teller and nitrogen configuration. To emulate restarting the initialization procedure, we jump to another random point along the trace, simulating the scrambling of the P1 pair Jahn-Teller axis and nitrogen spin state with the green laser.  

\begin{figure}
    \centering
    \includegraphics[width=0.5\linewidth]{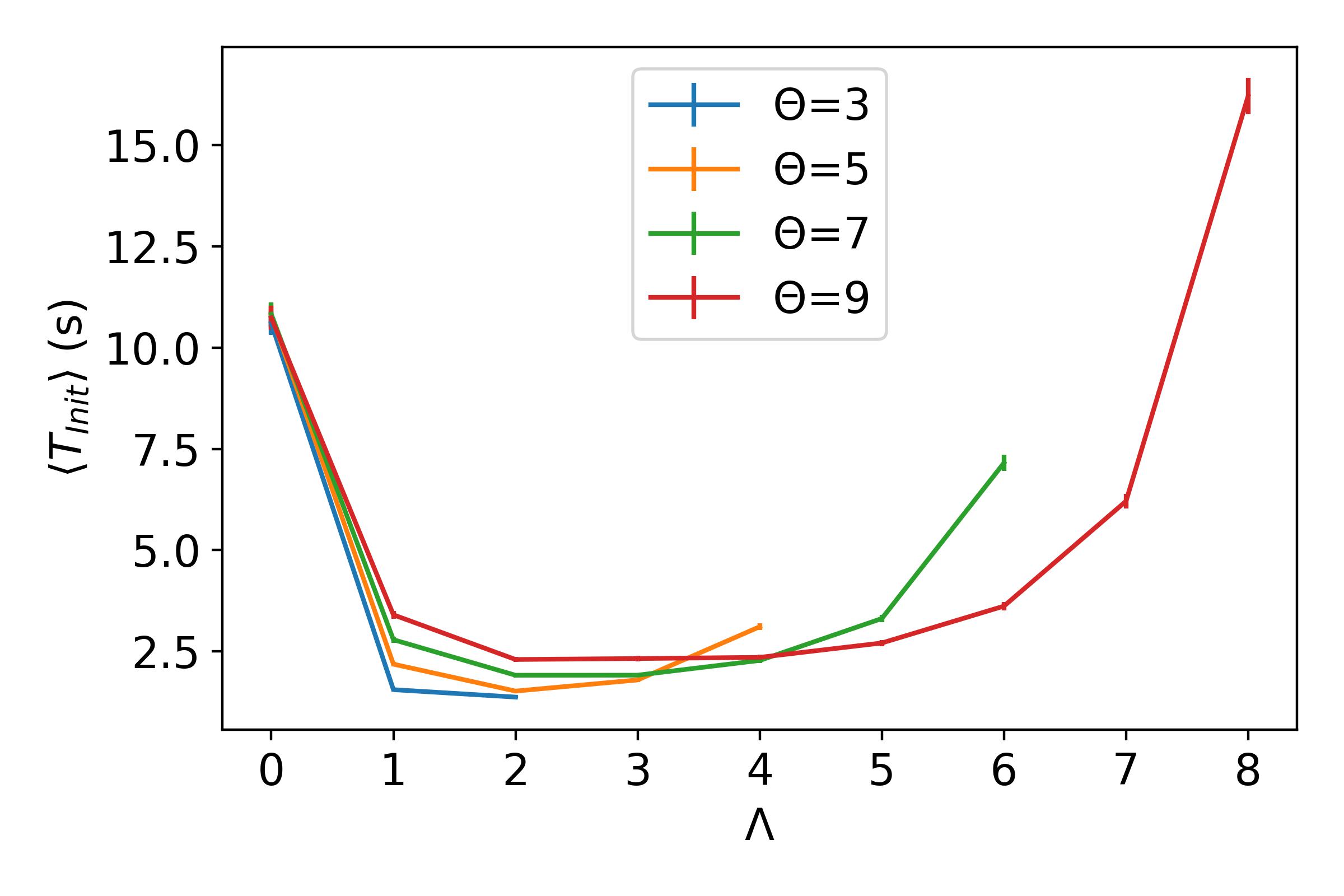}
    \caption{\textbf{Optimization of the initialization with a single threshold check.} We optimize the initialization time by sampling at different bin sizes $\Theta$ and sweep the allowed thresholds $\Lambda$. We find that the optimal combination is $\Theta = 3$ and $\Lambda = 2$, for which the average initialization time is $1.38(4)$s.}
    \label{fig:thresholds}
\end{figure}

We examine an initialization scheme with a single intermediate threshold check by sampling the data set for varying bin sizes $\Theta = {3, 5, 7, 9}$ and thresholds $\Lambda =[0,8]$. We find the minimum initialization time by sweeping over the possible thresholds $\Lambda$ for each size $\Theta$. As a figure of merit, we calculate the total time needed to achieve $5000$ successful P1 pair initializations. This is defined as both the intermediate and final check of $50$ parity outcomes surpassing their respective thresholds. The results are plotted in Fig. \ref{fig:thresholds}. We find an optimum set of parameters for a single check of $\Theta = 3$ and
$\Lambda = 2$. In this setting, the average time for each successful P1 pair initialization is $1.38(4)$ s. By introducing another intermediate check with bin size $\Theta = 10$ and threshold $\Lambda=3$ the initialization time slightly improves to $1.37(4)$ s. 

The final result provides a factor 10 improvement over the basic initialization without intermediate thresholds, essential to make the experiments in the main text feasible. We note that this is a crude optimization and that more advanced methods, like Bayesian inference or techniques based on machine learning, are likely to provide additional speed-ups. 

\begin{figure}
    \centering
    \includegraphics[width=0.8\linewidth]{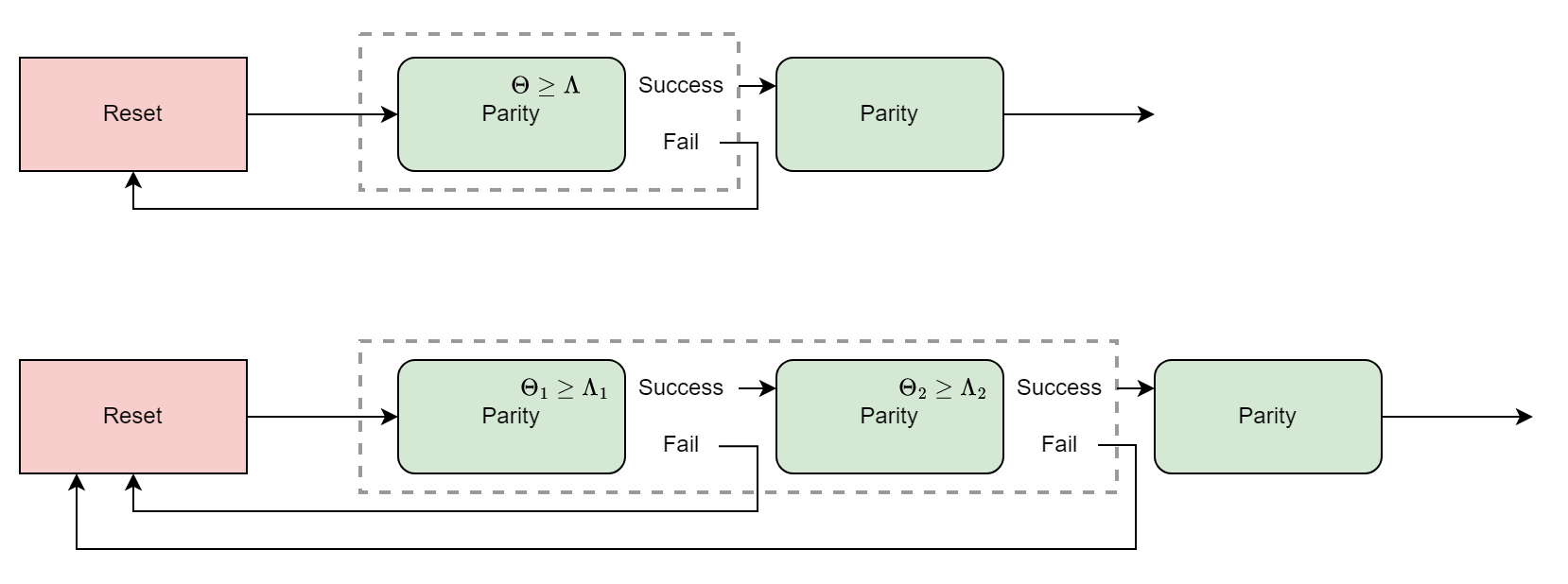}
    \caption{\textbf{Initialization schemes with single and double intermediate threshold checks.} For a successful initialization, we check for $15$ observed photons out of a total of $50$ parity measurements. However, we abort the procedure at one (top) or two (bottom) intermediate stages, depending on the photon counts thus far. With intermediate checks at $2$ out of $3$ photons, followed by $3$ out of $10$, we achieve an average initialization time of $1.37(4)$ s. }
    \label{fig:doublecheck}
\end{figure}

\clearpage

\section{Spin readout calibration} \label{spin_readout_calibration}

In this section, we describe the optimization of the spin readout. In particular, we show results for $\tau = 11.2$ \textmugreek s, but note that the procedure and results are similar for other values of $\tau$. The experiments typically consist of a two-step initialization process, containing first parity readouts and then spin readouts, and a one-step readout process, containing only spin readouts (Fig. 3). The optimization of the parity initialization has been described in detail in section \ref{optimization}. In the following, we will therefore assume that we aim to distinguish between the two anti-parallel states $\ket{\uparrow \downarrow}$ and $\ket{\downarrow \uparrow}$. 

For spin initialization, we do not perform any real-time thresholding. Passing the parity readout is a rare event (section \ref{optimization}) and it is therefore beneficial to collect all data. In the analysis, we then distinguish between the two pseudo-spin states by thresholding on the number of counts obtained in the spin readouts ($\geq 1/5$ or $0/5$).

\begin{figure}[h]
    \centering
    \includegraphics[width=\linewidth]{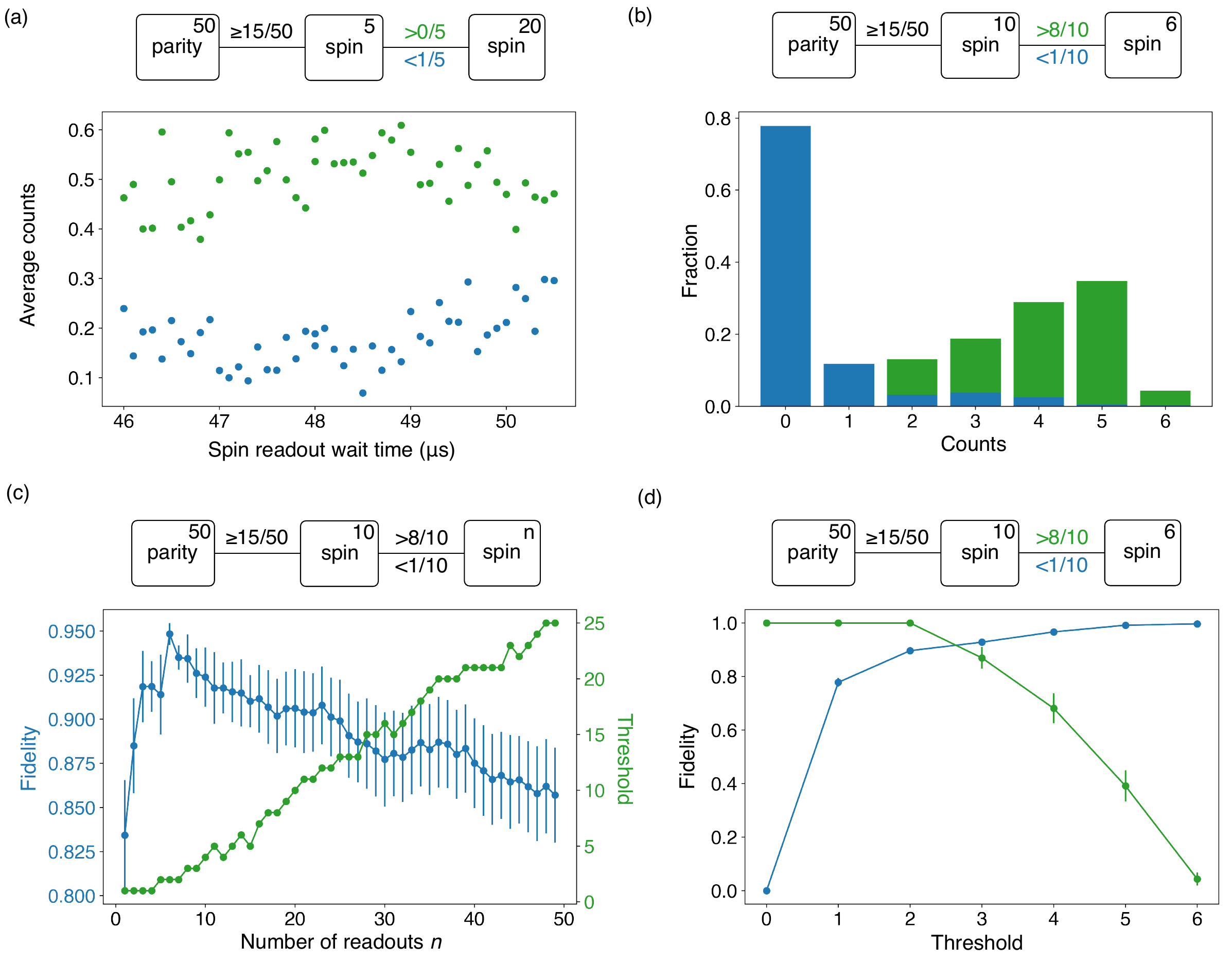}
    \caption{\textbf{Optimization of spin readout for $\tau = 11.2$ \textmugreek s.} \textbf{(a)} Sweep of the wait time between subsequent spin readouts. We find a maximum contrast between the two spin states around $48.5$ \textmugreek s. \textbf{(b)} Histogram for the two different spin states $\ket{\uparrow \downarrow}$ and $\ket{\downarrow \uparrow}$ for $n = 6$ spin readouts. \textbf{(c)} Sweep of the number of readouts $n$, showing the optimal fidelity $F$ and optimal threshold $T$ for each number of readouts. \textbf{(d)} Sweep of the threshold $T$ for $n = 6$ spin readouts. The optimal spin readout parameters are $n = 6$ readouts with a threshold of $T = 2$, giving a combined initialization and readout fidelity of $F = 94.8(6) \%$.}
    \label{fig:spin_readout}
\end{figure}

During a spin readout, the electron-spin pair evolves with frequency $X$, which is the dipolar coupling between the two spins of the spin pair. To make sure that each spin readout repetition measures along the same basis, we calibrate the time between subsequent spin readouts (Fig. \ref{fig:spin_readout}a). If the timing is correct, $\ket{\uparrow \downarrow}$ and $\ket{\downarrow \uparrow}$ will show a difference in obtained counts during the spin readout. The optimal timing is found around $48.5$ \textmugreek s. Note that the spin readout wait time has to be calibrated for each $\tau$ separately since the frequency $X$ is different for each $\tau$.

We perform the readout optimization as described in the supplementary material of Ref. \cite{Bartling2022}. We also discuss the process here for completeness. The two states that we want to optimally distinguish are $\ket{\uparrow \downarrow}$ and $\ket{\downarrow \uparrow}$, which we will write as $\ket{a}$ and $\ket{b}$ for simplicity. In the initialization step, we use $k$ repetitions and record $N(k)$ counts. We set strict thresholds to make the initialization as good as possible: $N(k) > N_a$ and $N(k) < N_b$ where $N_a$ ($N_b$) is the threshold to initialize in $\ket{a}$ ($\ket{b}$). In the readout step, we use $n$ repetitions and obtain $N(n)$ counts. To optimally distinguish states $\ket{a}$ and $\ket{b}$, we sweep a threshold $T$ and obtain the combined initialization and readout fidelity as:

\begin{equation}
    F = \frac{F_{\ket{a}} + F_{\ket{b}}}{2} = \frac{1}{2}P(N(n) \geq T | N(k) > N_a) + \frac{1}{2} P (N(n) < T | N(k) < N_b).
\end{equation}

We then optimize $F$ for the number of readouts $n$ and the threshold $T$. In Fig. \ref{fig:spin_readout}b, we show a histogram of the obtained counts for the two different spin states $\ket{\uparrow \downarrow}$ and $\ket{\downarrow \uparrow}$ for $n = 6$ spin readouts, obtained using strict initialization thresholds of $>8/10$ and $<1/10$. In Fig. \ref{fig:spin_readout}c, we sweep the number of spin readouts $n$ as well as the threshold $T$. We find an optimal number of readouts $n = 6$ with a threshold $T = 2$ obtaining a combined initialization and readout fidelity of $F = 94.8(6) \%$. In Fig. \ref{fig:spin_readout}d, we show a sweep of the threshold $T$ for $n = 6$ spin readouts, giving the optimal threshold of $2$.

For each value of $\tau$, we performed this characterization separately. However, the optimal parameters are very similar for other values of $\tau$. Hence, the number of spin readouts $n = 6$ and threshold $T = 2$ was used for all measured values of $\tau$. 

\clearpage

\bibliography{supp.bib}